\documentclass[fleqn,twoside]{article}
\usepackage{amsmath,amssymb,espcrc2,color}

\usepackage{graphicx,soul}
\usepackage[figuresright]{rotating}

\newcommand{\url}[1]{{\tt #1}}
\newcommand{\lsim}
{\;\raisebox{-.3em}{$\stackrel{\displaystyle <}{\sim}$}\;}

\newcommand{\gmt}{\ensuremath{(g-2)_\mu}}
\newcommand{\br}{{\rm BR}}

\newcommand{\bmm}{\ensuremath{\br(B_s \to \mu^+\mu^-)}}
\newcommand{\ssi}{\ensuremath{\sigma^{\rm SI}_p}}
\newcommand{\Och}{\ensuremath{\Omega_\chi h^2}}

\newcommand{\Mh}{M_h}

\newcommand{\mt}{m_t}
\newcommand{\mgl}{m_{\tilde g}}

\newcommand{\mneu}[1]{m_{\tilde \chi^0_{#1}}}

\newcommand{\tb}{\tan\beta}

\newcommand{\tev}{\,\, \mathrm{TeV}}
\newcommand{\gev}{\,\, \mathrm{GeV}}
\newcommand{\mev}{\,\, \mathrm{MeV}}
\newcommand{\cmsatlas}{CMS and ATLAS}

\definecolor{Orange}{named}{Orange}
\definecolor{Purple}{named}{Purple}

\newcommand{\ETslash}{/ \hspace{-.7em} E_T}

\graphicspath{{figs/}}

\title{\bf Implications of Initial LHC Searches for Supersymmetry
 \\ \vspace{0.5em}}

\author{
{\bf O.~Buchmueller}\address[Imperial]
   {High\,Energy\,Physics\,Group, Blackett\,Laboratory, Imperial\,College, 
    Prince\,Consort\,Road, London\,SW7\,2AZ,\,UK},
{\bf R.~Cavanaugh}\address[FNAL]
   {Fermi National Accelerator Laboratory, P.O. Box 500, 
    Batavia, Illinois 60510, USA}\hbox{$^{\rm ,}$}\address[UIC]
   {Physics Department, University of Illinois at Chicago, Chicago, 
    Illinois 60607-7059, USA},
{\bf D.~Colling}\addressmark[Imperial],
{\bf A.~De Roeck}\address[CERN]
   {CERN, CH--1211 Gen\`eve 23, Switzerland}\hbox{$^{\rm ,}$}\address[Antwerpen]
   {Antwerp University, B--2610 Wilrijk, Belgium},
 {\bf M.J.~Dolan}\address[IPPP]
{Institute for Particle Physics
     Phenomenology,\,University\,of\,Durham,\,South 
     Road,\,Durham\,DH1\,3LE,\,UK},
{\bf J.R.~Ellis}\addressmark[CERN]\hbox{$^{\rm ,}$}\address[KCL]{Theoretical Physics
  and Cosmology Group, Department of Physics, King's College London, London
  WC2R 2LS, UK}, 
{\bf H.~Fl\"acher}\address[Rochester]
   {Department of Physics and Astronomy, University of Rochester, 
    Rochester, New York 14627, USA},
{\bf S.~Heinemeyer}\address[Santander]
   {Instituto de F\'{\i}sica de Cantabria (CSIC-UC), 
    E--39005 Santander, Spain},
{\bf G.~Isidori}\address[Frascati]
   {INFN, Laboratori Nazionali di Frascati, Via E. Fermi 40, 
    I--00044 Frascati, Italy},
{\bf K.~Olive}\address[Minnesota] 
   {William\,I.\,Fine\,Theoretical\,Physics\,Institute,\,University\,of\,Minnesota,\,Minneapolis,\,Minnesota\,55455,\,USA}, 
{\bf S.~Rogerson}\addressmark[Imperial],
{\bf F.~Ronga}\address[ETHZ]
   {Institute for Particle Physics, ETH Z\"urich, CH--8093 Z\"urich, 
   Switzerland},
{\bf G.~Weiglein}\address[DESY]
   {DESY, Notkestrasse 85, D--22607 Hamburg, Germany}
}
       
\begin{document}

\begin{abstract}
The \cmsatlas\ Collaborations have recently published the 
results of initial direct LHC searches for supersymmetry analyzing
$~\sim 35$/pb of data taken at 7~TeV in the centre of mass. We
incorporate these results into a frequentist analysis of the probable
ranges of parameters of simple versions of the minimal supersymmetric
extension of the Standard Model (MSSM), namely the 
constrained MSSM (CMSSM), a model with common 
non-universal Higgs masses (NUHM1), the
very constrained MSSM (VCMSSM) and minimal supergravity (mSUGRA).
We present updated predictions for the gluino mass, $\mgl$, 
the light Higgs boson mass, $\Mh$, \bmm\ 
and the spin-independent dark matter scattering cross section, \ssi.
The \cmsatlas\ data make inroads into the CMSSM, NUHM1
and VCMSSM (but not mSUGRA) parameter spaces, thereby strengthening previous
lower limits on sparticle masses and upper limits on \ssi\ in the CMSSM and VCMSSM.
The favoured ranges of \bmm\ in the CMSSM, VCMSSM and mSUGRA are close to the
Standard Model, but considerably larger values of \bmm\ are possible in the NUHM1.
Applying the \cmsatlas\ constraints improves the consistency of the
model predictions for $\Mh$ with the LEP exclusion limits.

\bigskip
\begin{flushleft}
\vspace{-0.5cm}
\end{flushleft}
\begin{center}
\vspace{1.5cm}
{\tt CERN-PH-TH/2010-321, DCPT-11-02, DESY 10-249, IPPP-11-01, \\
FTPI-MINN-10/39, KCL-PH-TH/2011-04, UMN-TH-2933/10}
\end{center}
\vspace{-0.5cm}
\end{abstract}

\maketitle


The results of experiments at the LHC will be make-or-break for supersymmetry.
Multiple analyses have shown that the ATLAS and CMS experiments at the LHC
have excellent chances of discovering supersymmetry (SUSY)~\cite{ATLAS,CMS} if
it provides the astrophysical cold dark matter~\cite{EHNOS}, and/or if
sparticles are light enough to render natural the electroweak mass
scale~\cite{hierarchy}, and/or SUSY explains the apparent
discrepancy between experimental measurement and the Standard Model (SM)
prediction for the anomalous magnetic moment of the muon,
\gmt~\cite{newBNL,g-2,newDavier}. In parallel, the CDF, D\O, 
ATLAS and CMS experiments
should be able to establish or disprove the existence of a SM-like Higgs
boson weighing less than about $135\gev$, which is the upper bound
predicted by the minimal
supersymmetric extension  of the Standard Model
(MSSM)~\cite{erz,Degrassi:2002fi}, provided that SUSY is realized at the
TeV scale. 

In anticipation of the LHC start-up, many groups have ventured estimates
of the possible masses of supersymmetric particles in variants of the
MSSM~\cite{estimates}. The unconstrained MSSM contains too many 
parameters for a full exploration of its parameter space to be possible
using present data. Therefore, we have focused on making estimates
within the constrained MSSM (CMSSM)~\cite{cmssm1,mc1} in which soft
SUSY-breaking mass parameters are assumed to be universal at
the GUT scale, in the simplest generalization of this model in which the
universality is relaxed to allow non-universal Higgs masses
(NUHM1)~\cite{nuhm1,mc2,mc3}, in a very constrained model in which a
supplementary relation is imposed on trilinear and bilinear soft
SUSY-breaking masses (VCMSSM) \cite{vcmssm}, and in minimal supergravity
(mSUGRA) in which, in addition, the gravitino mass $m_{3/2}$ is set
equal to the common soft SUSY-breaking scalar  mass $m_0$
before renormalization~\cite{vcmssm,mc4}. 

Our estimates \cite{mc1,mc2,mc3,mc4,mc-web} have been made in
a  frequentist approach, in which we construct a global likelihood
function with contributions from precision electroweak observables,
$B$-physics observables, \gmt\ and the astrophysical cold dark matter density
\Och\ as well as the limits from the direct searches
for Higgs bosons and sparticles at LEP. 
Our best fits in the CMSSM, NUHM1, VCMSSM and
mSUGRA all suggested that sparticles should be relatively light, perhaps
even within the reach of early runs of the LHC. 

The first results of initial direct searches for SUSY in data recorded by the
\cmsatlas\ detectors analyzing $\sim 35$/pb of integrated luminosity of
collisions at 7~TeV in the centre of mass taken in 2010  have now
been published~\cite{CMSsusy,ATLASsusy}. In this paper we combine these
newly-published results with our previous analyses to give updated
predictions for the preferred regions of parameter space, gluino and
Higgs masses, \bmm\ and the spin-independent dark
matter scattering cross section $\ssi$ in the CMSSM, NUHM1, VCMSSM and mSUGRA
frameworks. 

We recall that the CMSSM~\cite{cmssm1} has four input parameters:  the
universal soft SUSY-breaking scalar and gaugino masses $(m_0, m_{1/2})$,
a universal trilinear soft SUSY-breaking parameter $A_0$ and the ratio of
Higgs v.e.v.'s $\tb$, as well as the sign of $\mu$ (the magnitude
of $\mu$ and the bilinear SUSY-breaking parameter $B_0$ are fixed by the
electroweak vacuum conditions). Concerning the latter, the results on
\gmt~\cite{newBNL,g-2,newDavier} strongly favor a positive sign.
In the NUHM1 \cite{nuhm1}, a common soft SUSY-breaking contribution to
the masses of the two Higgs doublets is allowed to vary independently,
so there are five independent parameters. On the 
other hand, the VCMSSM \cite{vcmssm} imposes the supplementary
constraint $B_0 = A_0 - m_0$ on the CMSSM, thereby removing $\tb$ as 
a free input and leaving three parameters, on which the further
constraint $m_{3/2} = m_0$ in mSUGRA~\cite{BIM} imposes a severe restriction. 

Our analysis has been performed using the 
{\tt MasterCode}~\cite{mc1,mc2,mc3,mc4,mc-web,mc35}. 
The model parameter spaces are sampled using a
Markov Chain Monte Carlo (MCMC) technique similar
to that used in our previous papers. 
Our MCMC samplings of the CMSSM and NUHM1 parameter spaces
each comprise some 25,000,000 points, whereas those of the VCMSSM and mSUGRA
include some 30,000,000 and 17,000,000 points, respectively.
The constraints are also treated similarly to our previous analyses, the
significant changes being an updated value of the top quark
mass, $m_t^{\rm exp} = 173.3 \pm 1.1 \gev$~\cite{mt1733} and the 
use of the new $e^+ e^-$ determination of the SM
contribution to \gmt~\cite{newDavier}, as described
in~\cite{mc4}~%
\footnote{We have not updated our evaluation of 
$R_{B\to\tau\nu} = {\rm BR}(B\to\tau\nu)/{\rm BR}(B\to\tau\nu)_{\rm SM} 
= 1.43 \pm 0.43$~\cite{mc3}, though higher values have been reported 
in~\cite{btn1,btn2}. All results are compatible within the errors: the 
main differences in the central values are
related to the values of $|V_{ub}|$ and $f_B$.
We have checked that using the value $R_{B\to\tau\nu} = 2.07 \pm 0.54$
given in~\cite{btn2}, corresponding to
less conservative treatments of $|V_{ub}|$ 
and $f_B$, would have negligible effects
on our fits, except to increase $\chi^2$ globally by $\sim 4$.
Also we note that a large non-SM contribution 
to $B\to\tau\nu$ is not supported by $K\to\mu\nu$ data~\cite{kmn}.}.
The numerical evaluation within the  
{\tt MasterCode}~\cite{mc1,mc2,mc3,mc4,mc-web,mc35},
combines
{\tt SoftSUSY}~\cite{Allanach:2001kg}, 
{\tt FeynHiggs}~\cite{Degrassi:2002fi,Heinemeyer:1998np,Heinemeyer:1998yj,Frank:2006yh},  
{\tt SuFla}~\cite{Isidori:2006pk,Isidori:2007jw},
{\tt SuperIso}~\cite{Mahmoudi:2008tp,Eriksson:2008cx}, 
a code providing supersymmetric predictions for electroweak observables based 
on~\cite{Heinemeyer:2006px,Heinemeyer:2007bw},
{\tt MicrOMEGAs}~\cite{Belanger:2006is,Belanger:2001fz,Belanger:2004yn} and
{\tt DarkSUSY}~\cite{Gondolo:2005we,Gondolo:2004sc}, making
extensive use of the SUSY Les Houches
Accord~\cite{Skands:2003cj,Allanach:2008qq}. 

The {\tt MasterCode} is designed in such a way that the constraints from
new observables can be taken into account and incorporated quickly and 
easily into the global likelihood function as `afterburners' 
(i.e., as add-ons to the global $\chi^2$ function), 
provided that the contribution to the likelihood function from the new
observable is available. 
The new ingredients in this analysis are the contributions from
the direct SUSY searches performed by \cmsatlas~\cite{CMSsusy,ATLASsusy},
which are incorporated as just such `afterburners', via the procedures
described below.

Previous studies~\cite{ATLAS,CMS} had indicated that $35$/pb at 
$7 \tev$ would provide sufficient sensitivity to probe regions 
of the $(m_0, m_{1/2})$ planes favoured previously at the 95\% and 68\%
confidence levels~(CL), close to the previous best-fit points in
the NUHM1 and CMSSM, though not as far as the best-fit points in the
VCMSSM and mSUGRA. Therefore, the possibility that the actual 
experimental exclusion would be less than 
the expected sensitivity  could not be excluded. However, this
amount of luminosity could not be expected to lead to a 5-$\sigma$
discovery of anomalous missing-transverse-energy events beyond those
expected in the SM, and the interpretation of any excess would be
ambiguous, since one could not expect to be able to discriminate between
SUSY and other potential explanations. On the other hand, it was to be
expected that even non-observation of SUSY with this event
sample would make an important contribution to the global likelihood
function and possibly alter significantly the results of the fits. 

The CMS result~\cite{CMSsusy} is based on a search for 
multijet + $\ETslash$ events without accompanying leptons. 
The 13 events found in the signal region were compatible with the $\sim 10.5$
expected from SM backgrounds with a probability value of 30\% 
The observed result allowed CMS to set a 95\%~CL (i.e.,
$1.96\,\sigma$) upper limit of 13.4 signal events. This would correspond 
to $2.5 \pm (13.4 - 2.5)/1.96 = 2.5 \pm 5.6$ events for any possible
signal, yielding $\chi^2_{\infty,{\rm CMS}} = 0.85$ for large sparticle masses.
The central CMS result, shown in Fig.~5 of ~\cite{CMSsusy}, is a 95\%~CL
exclusion contour in the $(m_0, m_{1/2})$ plane of the CMSSM for the
particular values $\tb = 3, A_0 = 0$ and $\mu > 0$. However, the
sensitivity of a search for multijet + $\ETslash$ events
is largely independent of these additional parameters within the
CMSSM~\cite{CMSsusy}, and can also be taken over to the NUHM1, VCMSSM and
mSUGRA models, which have similar signatures in these search channels.

Fig.~5 of~\cite{CMSsusy} also presents a $(m_0, m_{1/2})$ contour
for the 95\% CL exclusion expected in the absence of any signal, corresponding
to $5.56 \times 1.96 = 10.9$ events.  This contour 
would correspond to an {\it apparent} significance
of $(10.9 - 2.5)/5.56 \sim 1.5\,\sigma$ and hence $\Delta \chi^2 \sim 4$.
The observed 95\% CL contour, on the other hand,
corresponds to $\Delta \chi^2 = 5.99$.
We approximate the impact of the new CMS constraint by 
$\Delta \chi^2_{\rm CMS} \sim \chi^2_{\infty,{\rm CMS}} |(M_C/M) - 1|^{-p_C}$ 
(where $M \equiv \sqrt{m_0^2 + m_{1/2}^2}$) for each ray in the $(m_0, m_{1/2})$ 
plane, fitting the parameters $M_C, p_C$ by requiring 
$\Delta \chi^2 \sim 4, 5.99$ on the expected and observed 95\%
exclusion contours shown in Fig.~5 of~\cite{CMSsusy}~%
\footnote{We have checked that this procedure gives a value for 
$\Delta \chi^2$ at the LM1 point in Fig.~5 of~\cite{CMSsusy} that is
consistent with the 19.2 events expected. We note that the functional form
adopted for $\Delta \chi^2_{\rm CMS}$ is approximate, and our later 
estimates of systematic errors
in our results are based on studies of alternative forms.}.
  
The ATLAS result~\cite{ATLASsusy} is based on a search for 
multijet + $\ETslash$ events with one accompanying electron or muon. The 2
events found in the signal region were compatible with the $\sim 4.1$
expected from SM backgrounds with a probability value of 16\%.
The central ATLAS result, shown in Fig.~2 of ~\cite{ATLASsusy}, 
is again a 95\%~CL exclusion contour in
the $(m_0, m_{1/2})$ plane of the CMSSM for the
particular values $\tb = 3, A_0 = 0$ and $\mu > 0$, which is also
only moderately dependent on these additional parameters within the
CMSSM~\cite{ATLASsusy}, and can also be taken over to the NUHM1, VCMSSM and
mSUGRA models~%
\footnote{This would not be the case for multi-lepton + jets + $\ETslash$ 
searches, which have a stronger dependence on $\tb$, leading in
particular to weaker bounds for large $\tb$.}. 
The observed result allowed ATLAS to set a 95\%~CL (i.e.,
$1.96\,\sigma$) upper limit on sparticle production that corresponds to
4.8 signal events and $\chi^2 = 5.99$. This would correspond 
to a downward fluctuation of 
$-2.1 \pm (4.8 + 2.1)/1.96 = -2.1 \pm 3.5$ events for any possible
signal, yielding an estimate of $\chi^2_{\infty,{\rm ATLAS}} = 1.2$
for large sparticle masses.

In order to estimate the ATLAS sensitivity to sparticle masses 
in the parameter region beyond the observed 95\%~CL,
we follow~\cite{ATLASnote}, based on~\cite{ATLASstudy}, in which the
integrated luminosities ${\cal L}$ required for discovering sparticles of various
masses were estimated. There it was found empirically that ${\cal M}
\propto {\cal L}^{1/4}$, where ${\cal M}$ denotes a (similar)
squark and gluino mass.
This suggests that for equal squark and gluino masses
the effective event rate $\propto {\cal M}^{-4}$. Using this as a guide,
we assume that the effective numbers 
of events expected for relevant points in the $(m_0, m_{1/2})$ plane
scale as ${\cal M}^{-4}$. We then calculate the corresponding significances
assuming the observed signal of $-2.1 \pm 3.5$ estimated above, 
to estimate the corresponding values of $\chi^2$.
As an approximate analytic interpolating form, we use
$\Delta \chi^2_{\rm ATLAS} \sim \chi^2_{\infty,{\rm ATLAS}} + (M_A/M)^{-p_A}$, fitting $M_A$ and $p_A$
along rays in the $(m_0, m_{1/2})$ plane.
Our analysis is based exclusively on published material~\cite{CMSsusy,ATLASsusy,ATLASnote,ATLASstudy}, 
and we consider it prudent to assign a larger
systematic error to our implementation of the ATLAS constraint.
  
As shown in Fig.~5 of~\cite{CMSsusy} and Fig.~2 of~\cite{ATLASsusy}, the
direct physics reaches of the CMS and ATLAS data in the 
models studied are much greater than those of the earlier CDF, D\O\ and LEP2
searches for sparticles, so the latter do not make significant
contributions to the global $\chi^2$ function in the regions of interest
for our analysis.  
On the other hand, the LEP2 Higgs search~\cite{Barate:2003sz,Schael:2006cr}
had an indirect reach in the $(m_0, m_{1/2})$ plane that is comparable to
those of the direct CMS and ATLAS constraints in the models studied, as we see
explicitly later. 

\begin{figure*}[htb!]
\resizebox{8cm}{!}{\includegraphics{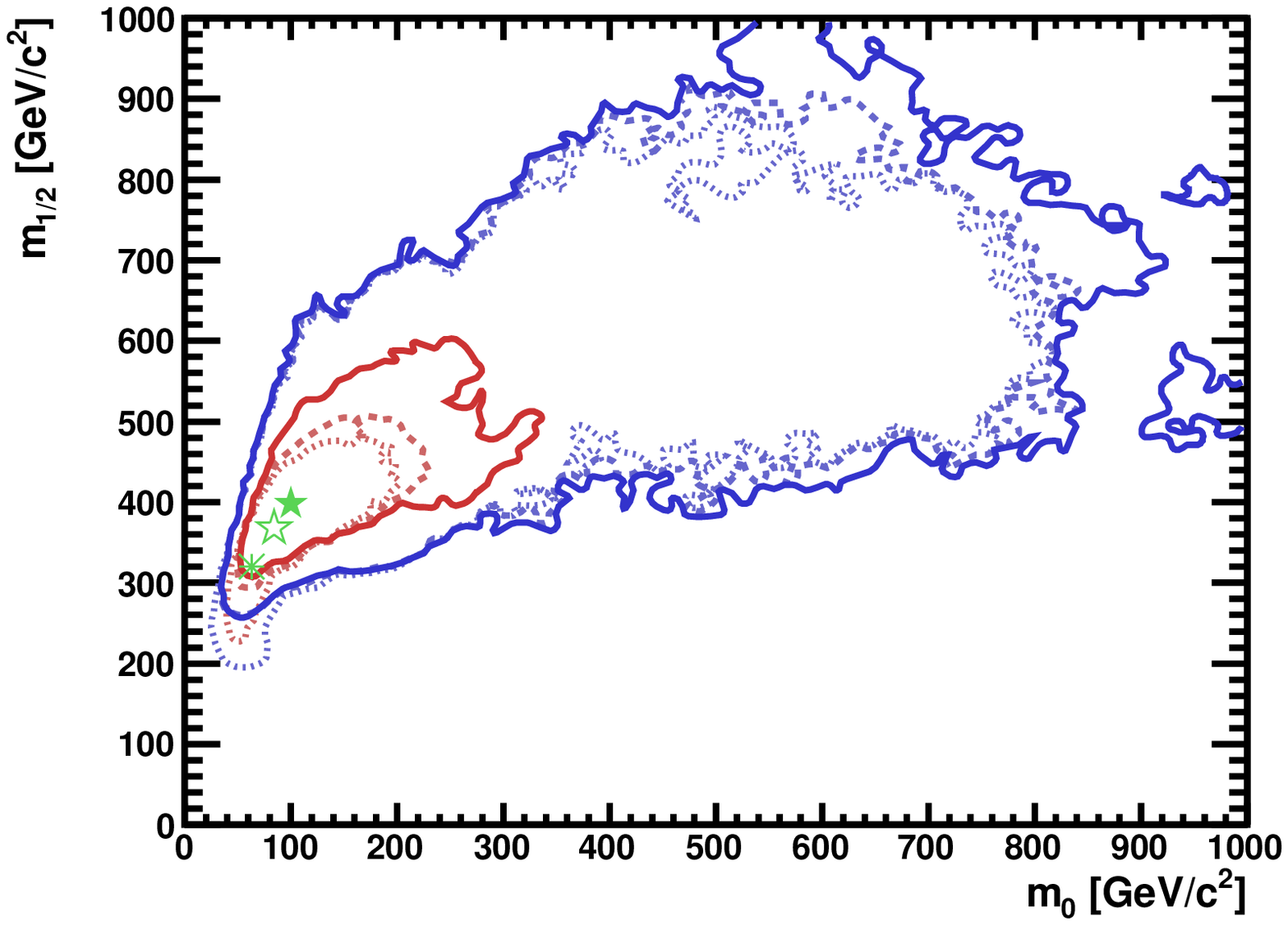}}
\resizebox{8cm}{!}{\includegraphics{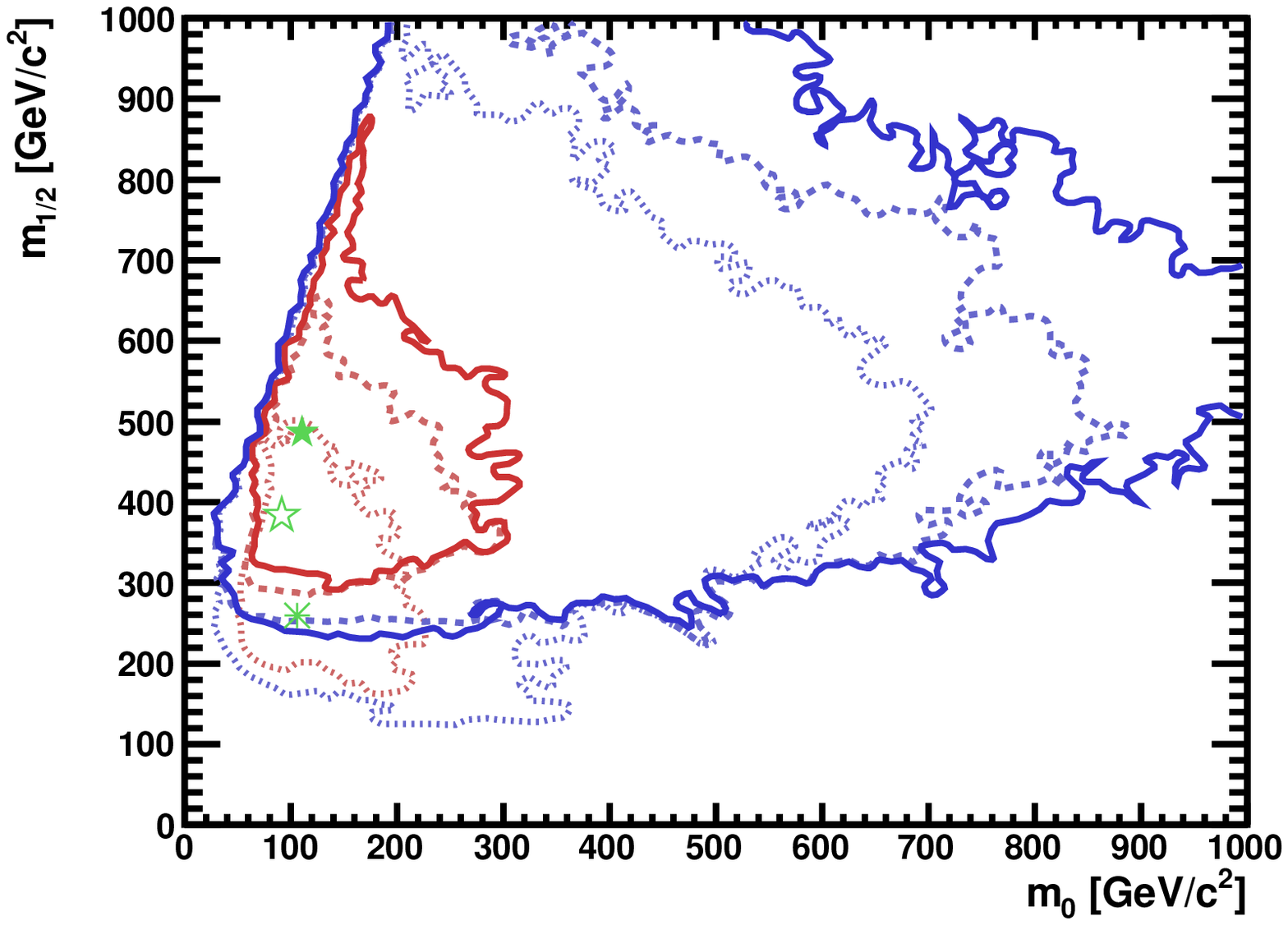}}
\resizebox{8cm}{!}{\includegraphics{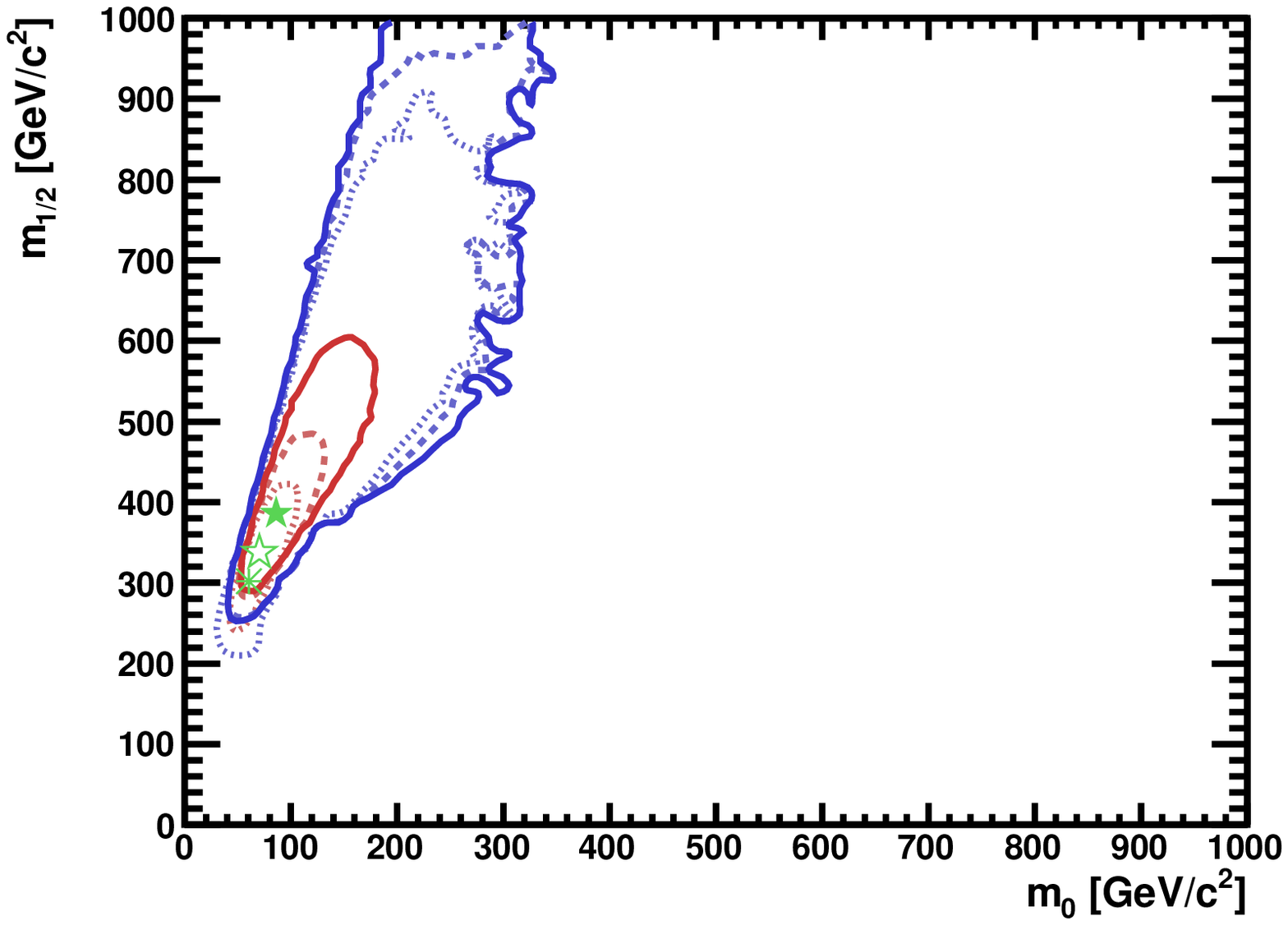}}
\resizebox{8cm}{!}{\includegraphics{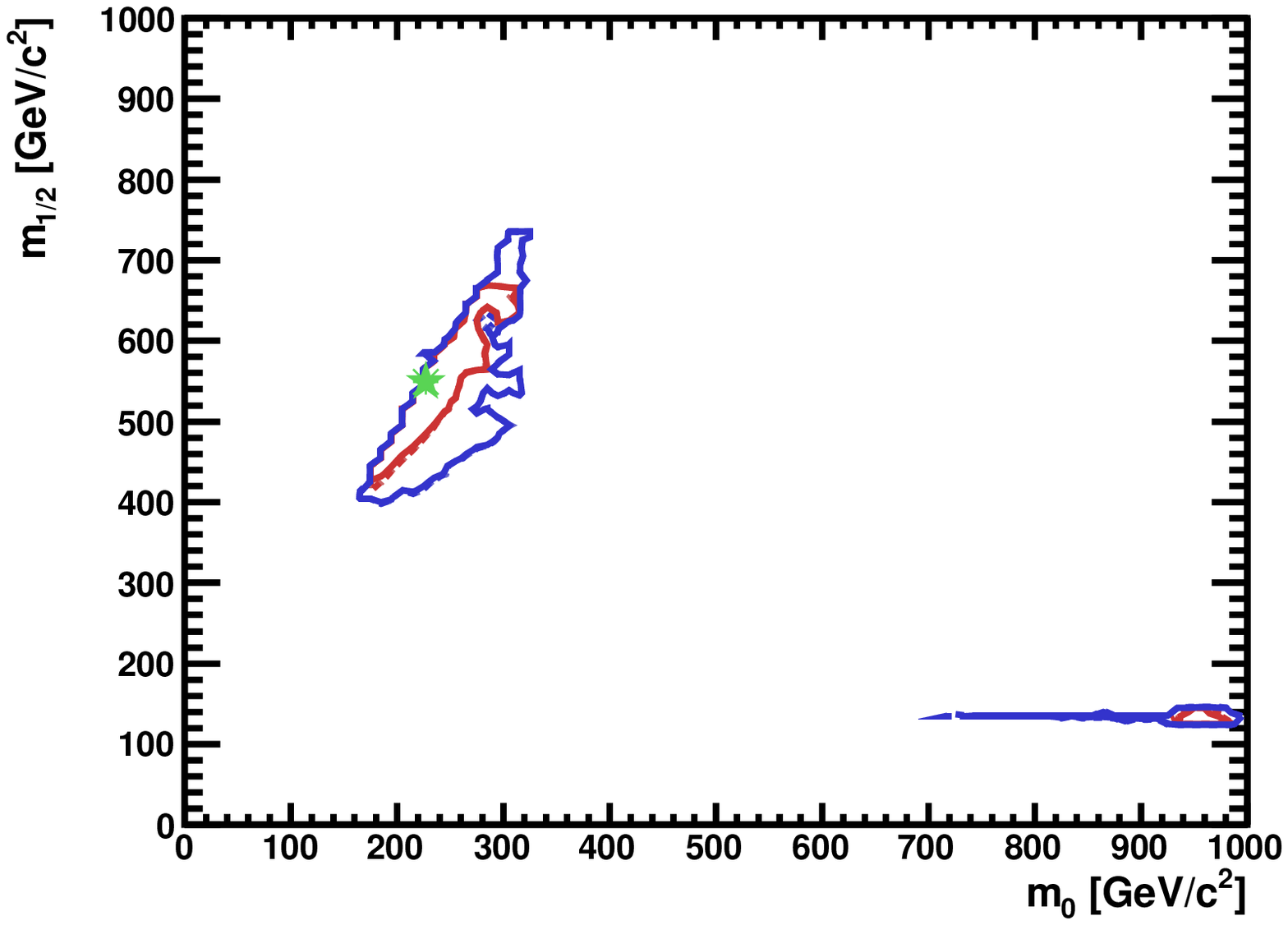}}\\
\vspace{-1cm}
\caption{\it The $(m_0, m_{1/2})$ planes in the CMSSM (upper left),
NUHM1 (top right), VCMSSM (lower left) and mSUGRA (lower right).
In each panel, we show  the 68 and 95\%~CL contours (red and blue,
respectively) both after applying the
CMS~\protect\cite{CMSsusy} and ATLAS~\protect\cite{ATLASsusy}
constraints (dashed and solid lines, 
respectively) and beforehand (dotted lines). Also shown as open
(solid) green stars are the best-fit points found after applying the 
CMS (ATLAS) constraints in each model (see text), and as green
`snowflakes' the previous best-fit points.}
\label{fig:m0m12}
\end{figure*}

Conservatively, we do not attempt to 
combine the CMS and ATLAS constraints in the
following discussion, which would require detailed modelling of their
likelihood functions and a better understanding of
correlations between observables in the CMS and ATLAS searches. 
The $\chi^2$ functions obtained by combining the CMS and
ATLAS constraints separately with previous constraints 
in the $(m_0, m_{1/2})$ planes for the CMSSM, NUHM1, VCMSSM and mSUGRA models
are shown in Fig.~\ref{fig:m0m12}. In each panel, the new 68\% and 95\%~CL
contours incorporating the CMS (ATLAS) results are shown as red and blue
dashed (solid) lines, and  the corresponding previous contours are shown as
dotted lines. (It should be noted that the updated values of $\mt$
and \gmt\ and some further technical improvements have changed slightly
the best fits and likelihood contours with respect 
to our earlier publications~\cite{mc2,mc3,mc4}.)
We also indicate as open and full green
stars the new best-fit points including
the \cmsatlas\ results, respectively, and indicating the pre-LHC
best-fit points by green `snowflakes'~%
\footnote{We recall
that in mSUGRA there is a secondary minimum of the likelihood function with
$\Delta \chi^2 \sim 4$ that is located at 
$(m_0, m_{1/2}) \sim (2110, 130) \gev$~\cite{mc4}, which is
not visible in Fig.~\ref{fig:m0m12} and lies well beyond the initial
LHC reach.}. 
In panels where only one green star is visible, the CMS and ATLAS
best-fit points lie on top of each other, within the $\sim 10 \gev$
accuracy adopted for our numbers.
The jagged boundaries of the 95\% CL CMSSM and NUHM1 contours
at large $m_0$ and $m_{1/2}$ reflect the uncertainties in sampling
the slow variations in their likelihood functions.

\begin{table*}[!tbh]
\renewcommand{\arraystretch}{1.2}
\begin{center}
\begin{tabular}{|c||c|c|c|c|c|c||c|c|} \hline
Model & Minimum $\chi^2$ & Probability & $m_{1/2}$ & $m_0$ & $A_0$ & $\tb$ & $\Mh$ (no LEP) \\
& & & (GeV) & (GeV) & (GeV) & & (GeV)\\ \hline \hline
CMSSM      & (21.3) & (32\%) & (320) & (60) & (-170) & (11) & (107.9) \\
with CMS   &  22.0  &   29\%    &  370  &  80  &   -340  &  14  &   112.6 \\
with ATLAS &  24.9  &   16\%    &  400  & 100  &   -430  &  16  &   112.8 \\
\hline
NUHM1      & (19.3) & (31\%) & (260) & (110) & (1010) & (8) & (121.9) \\
with CMS   &  20.9  &   28\%    &  380  &   90  &   70  & 14  &  113.5 \\
with ATLAS &  23.3  &  18\%    &  490  &  110  &   -630  & 25  &  116.5 \\
\hline
VCMSSM     & (22.5) & (31\%) & (300) &  (60) &   (30) & (9) & (109.3) \\
with CMS   &  23.8  &   25\%    &  340  &   70  &   50  &  9  &  115.5 \\
with ATLAS &  27.1  &   13\%    &  390  &   90  &   70  & 11  &  117.0 \\
\hline
mSUGRA     & (29.4) & (6.1\%) & (550) & (230)  & (430) & (28) & (107.8) \\
with CMS   &   29.4  &     6.1\%    & 550  & 230 & 430 & 28 &  121.2 \\
with ATLAS &  30.9  &   5.7\%      & 550  & 230 & 430 & 28 &  121.2 \\
\hline
\end{tabular}
\caption{\it Comparison of the best-fit points found previously in the CMSSM, the NUHM1,
the VCMSSM and the coannihilation region of mSUGRA when the LHC constraints were
not included (in parentheses)~\cite{mc2,mc3,mc4}, and the results of this
paper incorporating the CMS~\protect\cite{CMSsusy} and ATLAS~\protect\cite{ATLASsusy}
constraints. In addition to the minimum
value of $\chi^2$ and the fit probability in each scenario, we include the
values of $m_{1/2}, m_0, A_0$ and $\tb$ at  all the best-fit points, as well
as the predictions for $\Mh$ {\it neglecting} the LEP constraint.} 
\label{tab:compare}
\end{center}
\end{table*}

In Table~\ref{tab:compare} we compare the best-fit points
found in this paper incorporating the CMS and ATLAS constraints with pre-LHC
results~\cite{mc2,mc3,mc4} in the CMSSM, NUHM1 and VCMSSM
(the fits in mSUGRA are essentially unchanged when LHC data are included in
the fits, only the best coannihilation fit is reported). In addition to
the minimum value of $\chi^2$ and the fit 
probability in each scenario, we include the values of $m_{1/2}, m_0, A_0$ and
$\tb$ at  all the best-fit points. We estimate systematic errors of
$\sim 10\%$ in the values for $m_{1/2}$ quoted in
Table~\ref{tab:compare} for the best-fit points 
in the CMS analyses of the CMSSM, NUHM1 and VCMSSM, associated with the
ambiguities in the implementations of the LHC constraints and the slow
variations in the $\chi^2$ functions. In the cases of the ATLAS analyses, we have an additional
systematic uncertainty in the implementation of the constraint, and estimate a somewhat larger
error $\sim 20\%$.
Table~\ref{tab:compare} also shows the values of $\Mh$ that would be
estimated in each model if the LEP Higgs constraint were neglected.

The absences of supersymmetric signals in the
\cmsatlas\ data~\cite{CMSsusy,ATLASsusy} invalidate portions of the CMSSM,
NUHM1 and VCMSSM parameter spaces at low $m_{1/2}$ that were previously
allowed at the 95\% and 68\%~CLs~%
\footnote{Strictly speaking, the $\Delta \chi^2 = 5.99$ and 2.30 contours that we plot
may not always correspond exactly to these CLs, but we ignore any possible differences here
since previous studies~\cite{mc2,mc3,mc4} showed acceptable coverage, which should 
not have deteriorated in the presence of additional constraints. For recent studies in a
related context, see~\cite{oops}.}, 
but do not impinge significantly on the corresponding regions for
mSUGRA. In the cases of the CMSSM 
and VCMSSM, the LHC data disfavour the
low-$m_{1/2}$ tips of the coannihilation regions and increase significantly
the best-fit values of $m_{1/2}$, 
as seen in Fig.~\ref{fig:m0m12} and Table~\ref{tab:compare}.
However, it should be kept in mind that $\chi^2$ is quite a shallow
function of $m_{1/2}$ near the best-fit points, particularly in the NUHM1.
In the case of the NUHM1, the \cmsatlas\ data disfavour a slice of
parameter space at low $m_{1/2}$ and $m_0 < 400 \gev$ extending from the
coannihilation region towards the light-Higgs funnel discussed in~\cite{mc4}.
The CMS and particularly ATLAS data extend the 95\% CL regions to larger $m_0$
and $m_{1/2}$, particularly in the NUHM1.

While the CMS constraint, as seen in Table~\ref{tab:compare},
leads only to small increases in the global $\chi^2$ of $\sim 1$ for each model, 
the inclusion of the ATLAS constraint results in increases in
$\chi^2$ by $\lsim 4.6$.
This indicates that there is no tension between the CMS data and
previous constraints, whereas some tension may arise from the ATLAS constraint.
Correspondingly, the fit probabilities in the different models are reduced by including
the CMS and ATLAS constraints, but generally not to unacceptable levels.
It should be recalled, though, that our ATLAS implementation is more uncertain, and our
implementations of the constraints should, conservatively, each be assigned an uncertainty at least as large as the value of $\chi^2_\infty = 0.85, 1.2$ for CMS and ATLAS, respectively.
On the other hand, the absence of a supersymmetric signal at the LHC with a
luminosity of $\ge 1$/fb 
at a centre-of-mass energy $\ge 7$ TeV would increase
the global minimum of $\chi^2$ sufficiently to put severe pressure on
these models.

\begin{figure*}[htb!]
\resizebox{8cm}{!}{\includegraphics{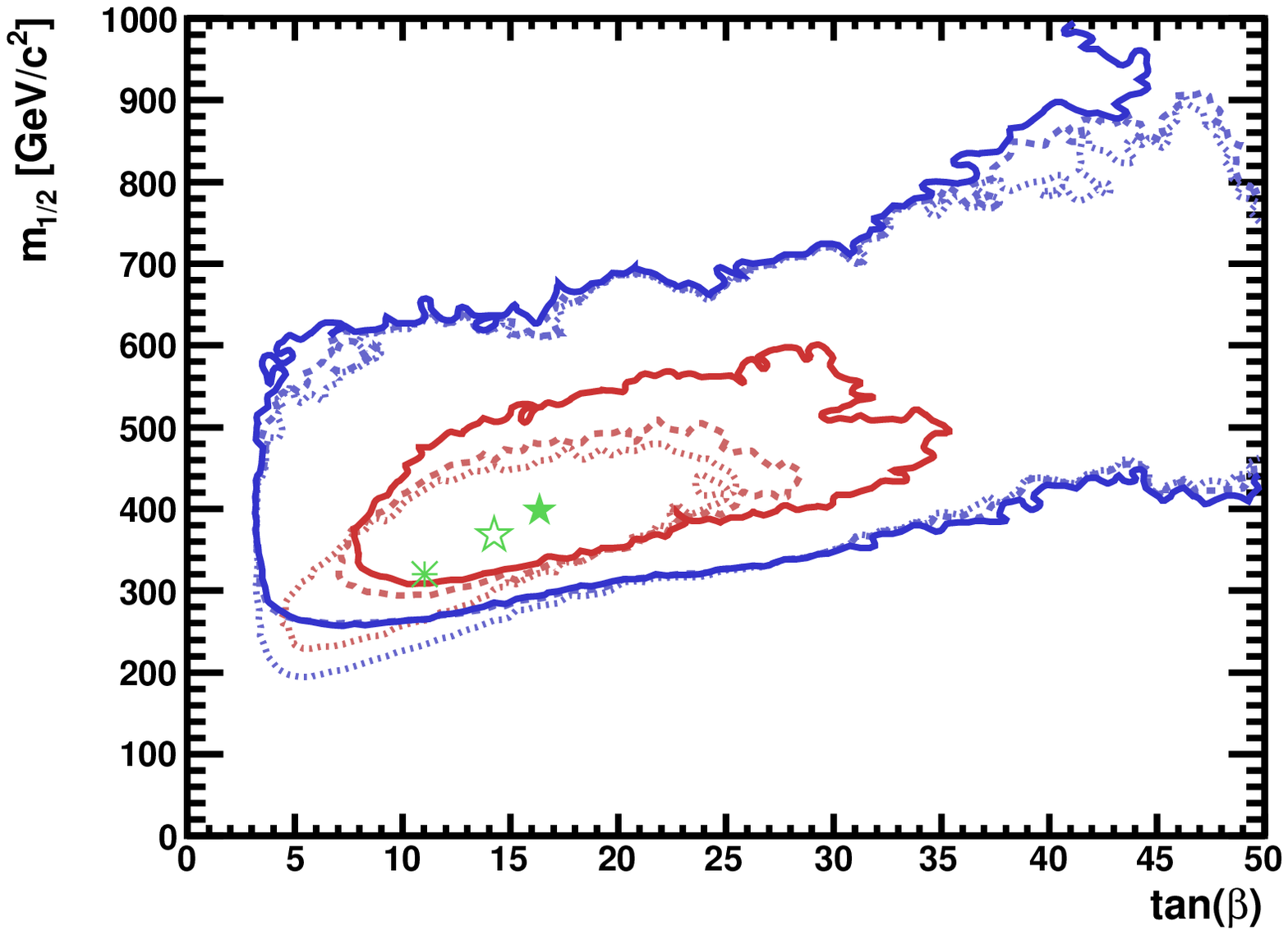}}
\resizebox{8cm}{!}{\includegraphics{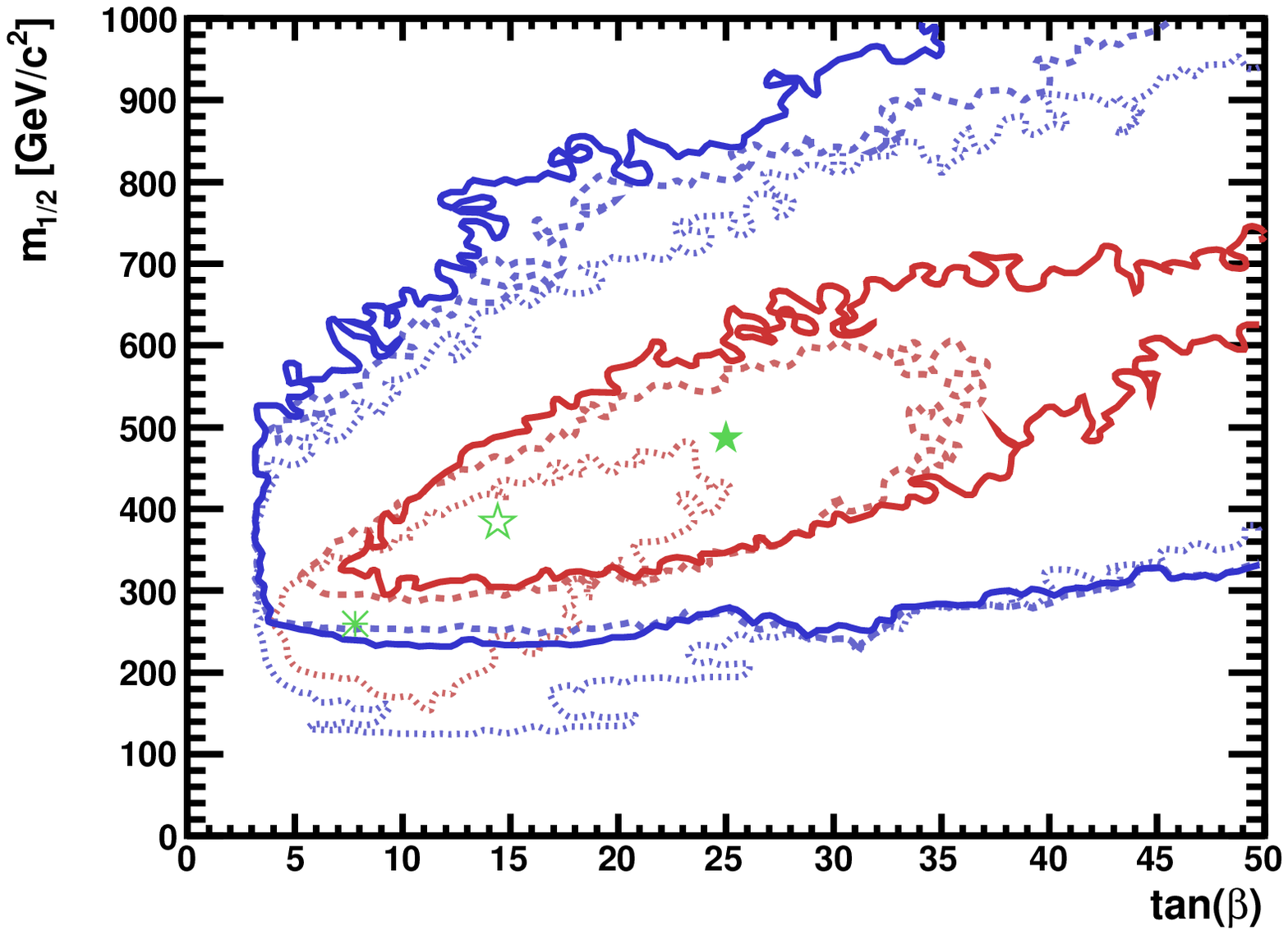}}
\resizebox{8cm}{!}{\includegraphics{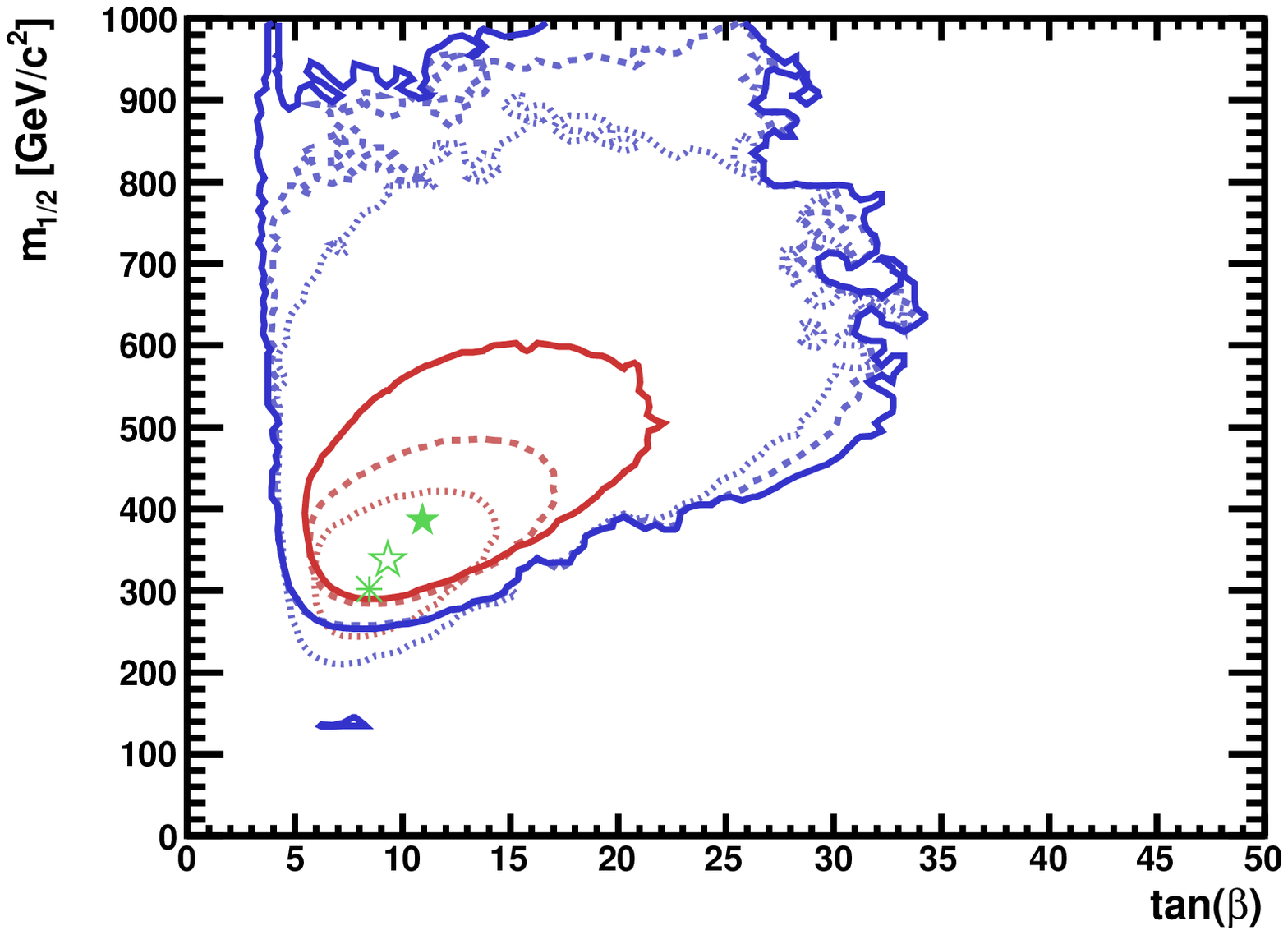}}
\resizebox{8cm}{!}{\includegraphics{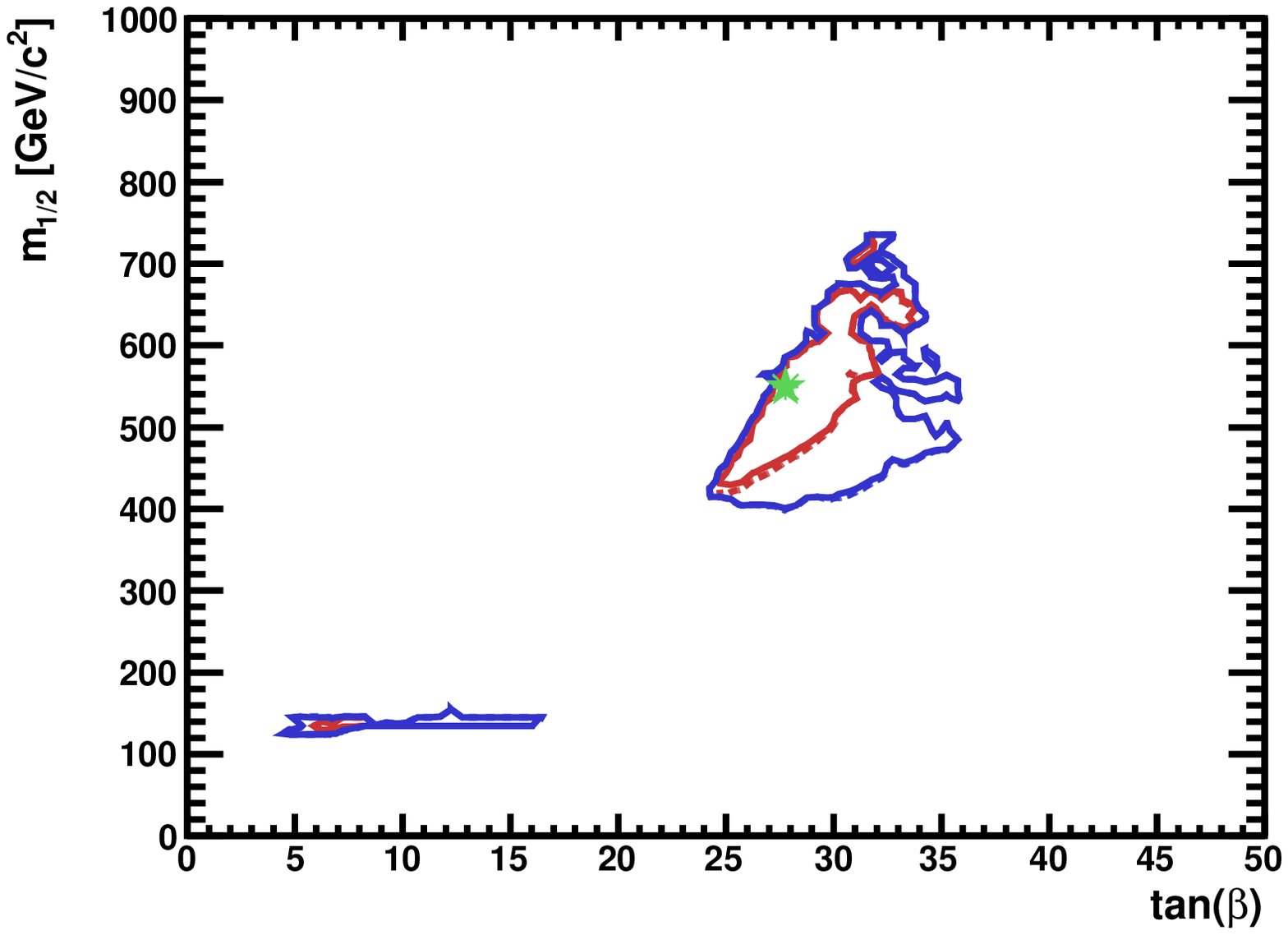}}\\
\vspace{-1cm}
\caption{\it The $(\tb, m_{1/2})$ planes in the CMSSM (upper left),
NUHM1 (top right), VCMSSM (lower left) and mSUGRA (lower right).
The line styles and stars have the same meanings as in
Fig.~\protect{\ref{fig:m0m12}}.} 
\label{fig:tBm12}
\end{figure*}

Fig.~\ref{fig:tBm12} displays the effects of the CMS and ATLAS
constraints on the $(\tb, m_{1/2})$ planes in the CMSSM, NUHM1, VCMSSM
and mSUGRA~%
\footnote{These planes are truncated at $\tan \beta = 50$, so that
perturbative RGE calculations remain reliable.}%
. We see that the ranges of $\tb$ preferred at the 68\% CL are
extended to significantly larger values in the CMSSM and (particularly)
the NUHM1, to a lesser extent in the VCMSSM, and not at all in
mSUGRA. The best-fit values of $\tb$ are also increased in the
CMSSM and NUHM1, but not significantly in the VCMSSM and mSUGRA.
The increases are mainly due to the $(g-2)_\mu$ constraint: larger sparticle masses
lead to a smaller supersymmetric contribution that can be compensated by a larger
value of $\tb$.

\begin{figure*}[htb!]
\resizebox{8cm}{!}{\includegraphics{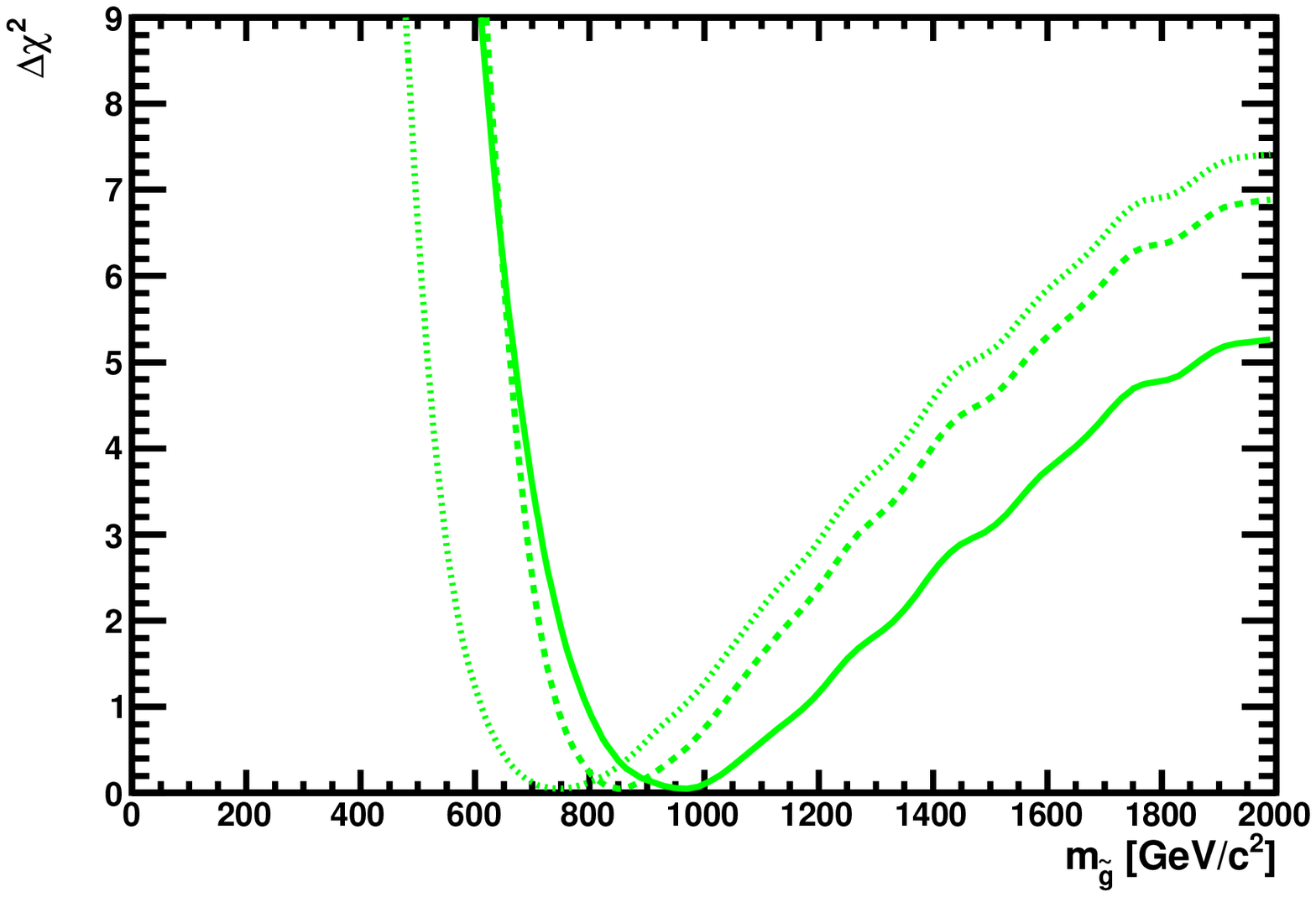}}
\resizebox{8cm}{!}{\includegraphics{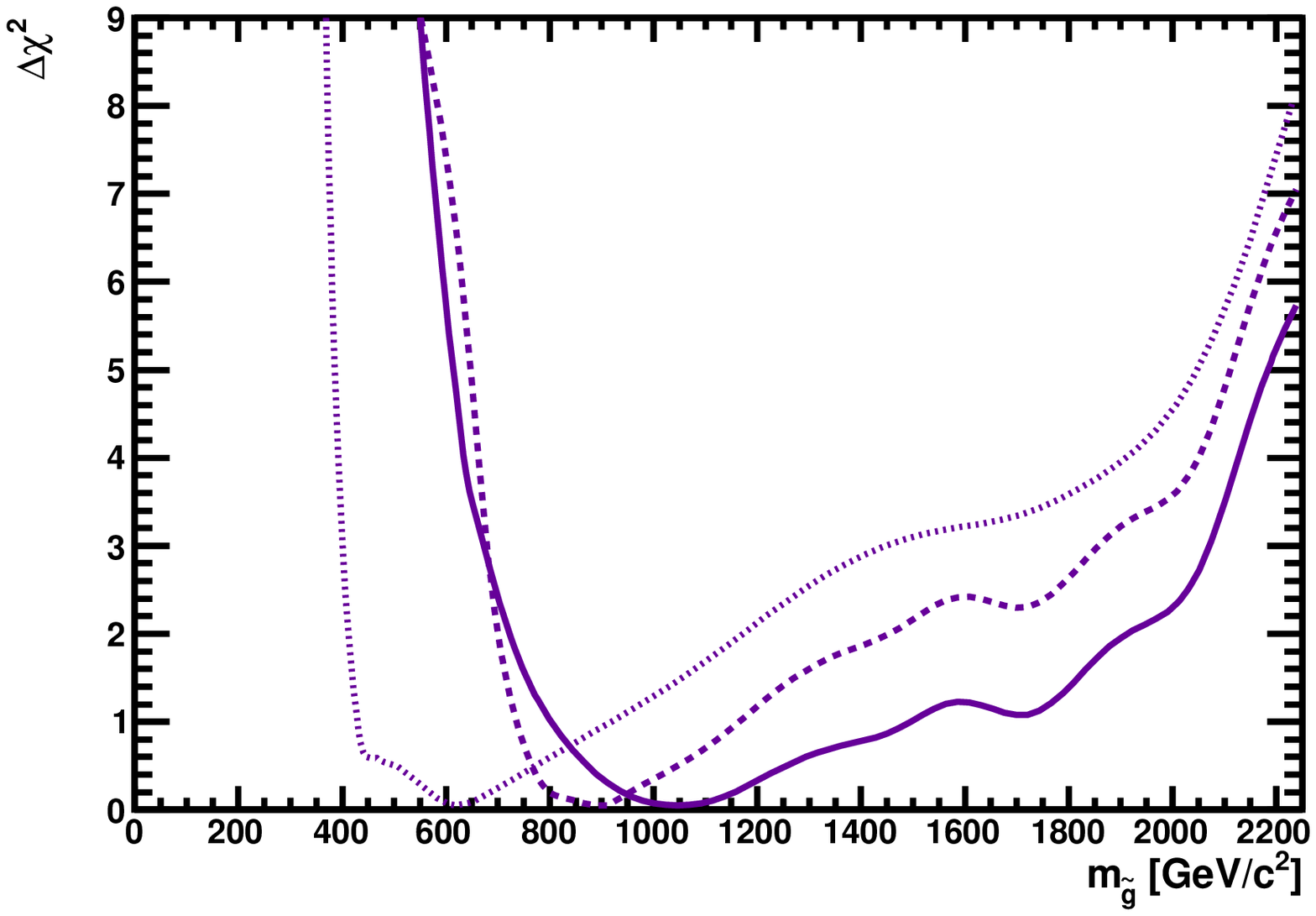}}
\resizebox{8cm}{!}{\includegraphics{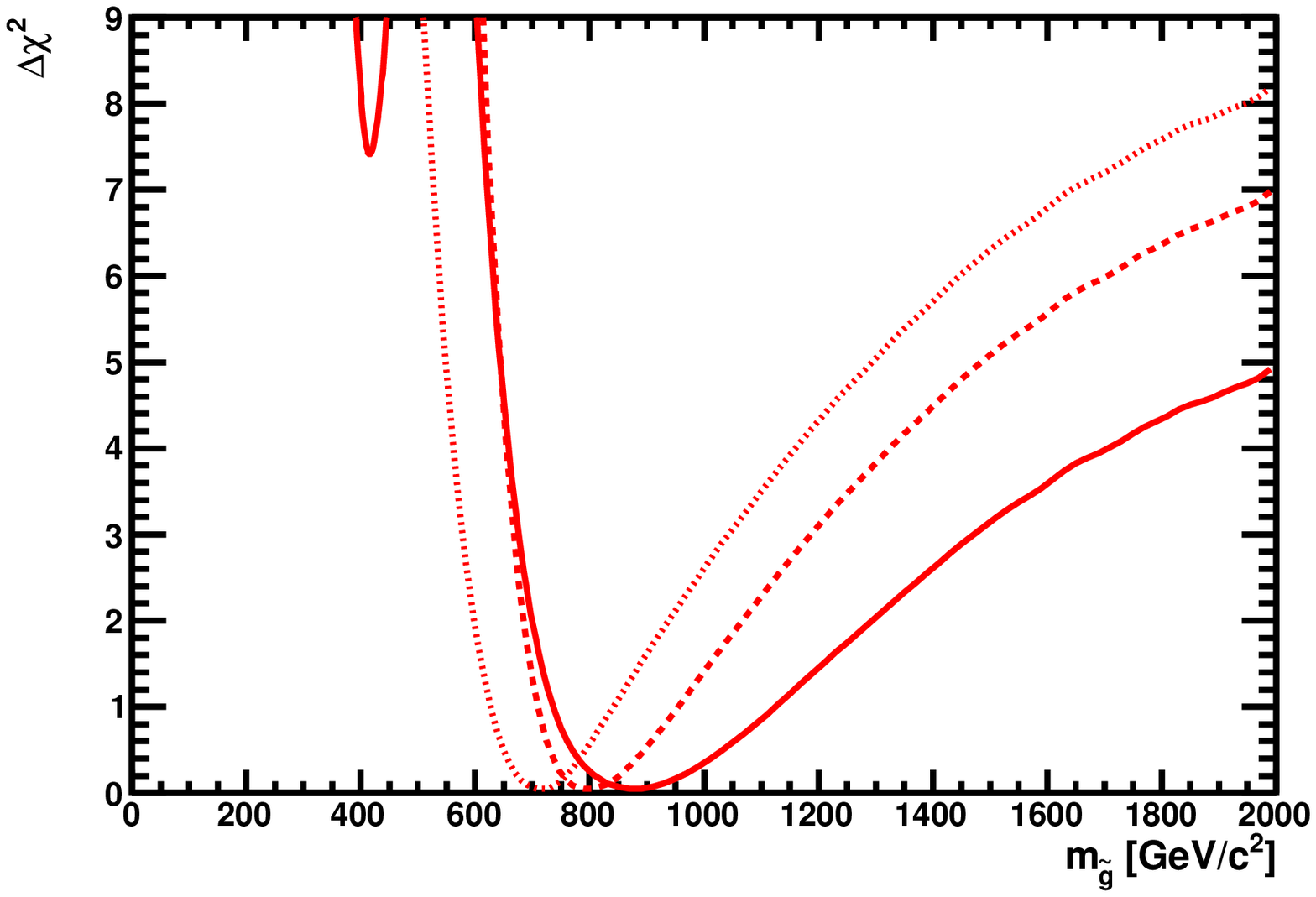}}
\resizebox{8cm}{!}{\includegraphics{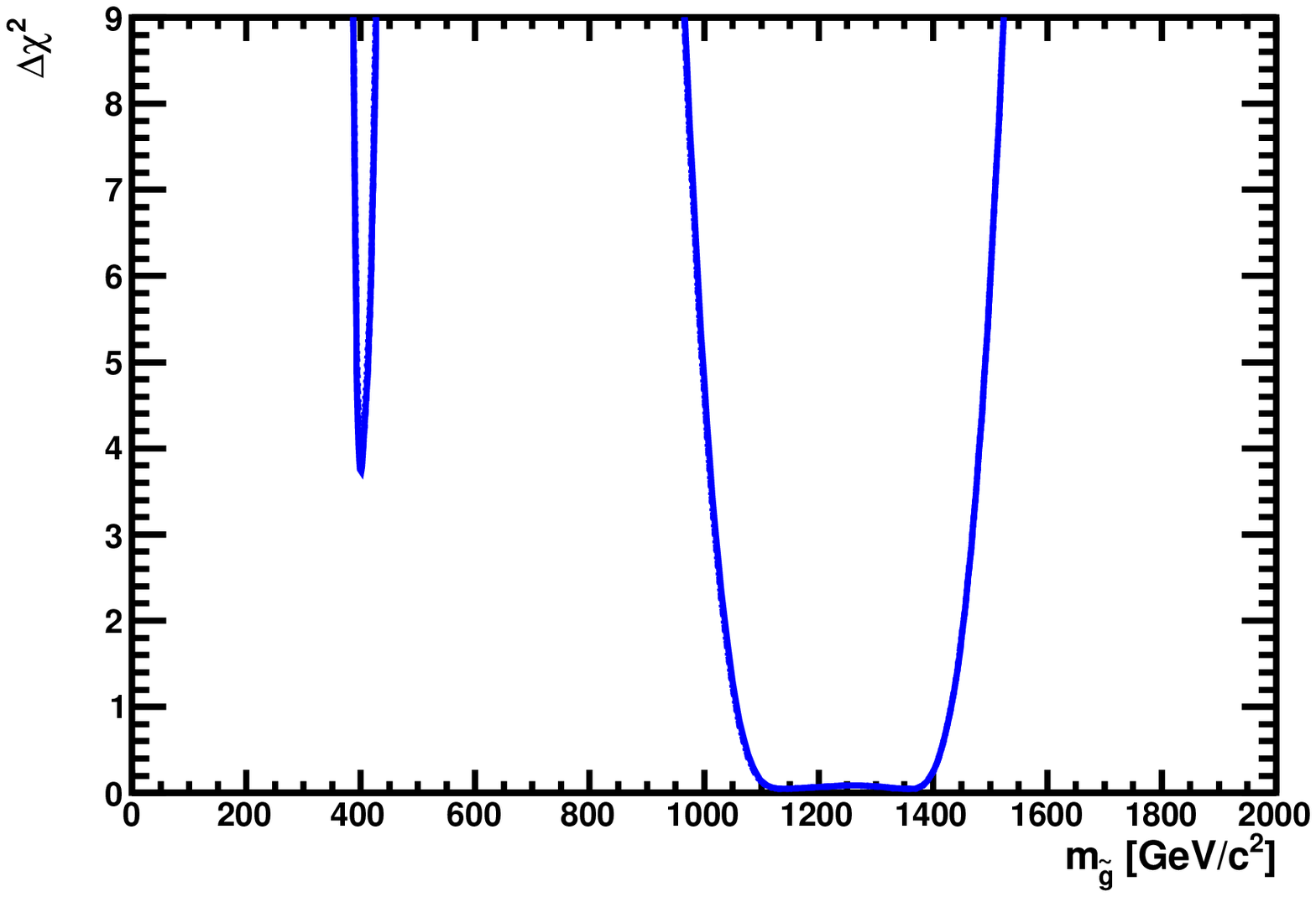}}\\
\vspace{-1cm}
\caption{\it The one-parameter $\chi^2$ likelihood functions for the gluino
mass $\mgl$ in the CMSSM (upper left),
NUHM1 (top right), VCMSSM (lower left) and mSUGRA (lower right).
In each panel, we show  the $\chi^2$ function including the
\cmsatlas~\protect\cite{CMSsusy,ATLASsusy} 
constraints as dashed and solid lines, respectively, and the previous
$\chi^2$ function as a dotted line.}
\label{fig:mgl}
\end{figure*}

In the regions of parameter space of interest to the CMSSM, NUHM1, VCMSSM
and mSUGRA, the direct reach of the LHC for supersymmetry is
strongly influenced by the 
gluino mass, $\mgl$. Accordingly, we display in Fig.~\ref{fig:mgl} the
one-parameter $\chi^2$ functions for $\mgl$ relative to the minima in
all these models. In each  
case, we display the new likelihood functions incorporating \cmsatlas\
data as dashed and solid lines, respectively, and those given by the
pre-LHC fits as dotted lines. The plots display the $\Delta\chi^2$
contributions in each model relative to the best-fit points in that
model.
In this and subsequent figures, the one-parameter 
$\chi^2$ functions for mSUGRA are essentially unchanged when the LHC data are
included but are shown for comparison purposes.

For each of the CMSSM, NUHM1 and VCMSSM, we see that the side of
the likelihood function below the best-fit point is shifted to larger $\mgl$
by similar amounts $\delta \mgl \sim 100$ to $400 \gev$.
The best-fit values of $\mgl$ in the CMSSM, NUHM1 and VCMSSM are now
$\sim 800$ to $1000 \gev$, and the sides of the likelihood function
beyond the best-fit points rise quite similarly in these models, though more
slowly in the NUHM1.
In mSUGRA the most likely values of $\mgl$ are unchanged by either CMS or ATLAS, lying in the range
$\sim 1100$ to $\sim 1400 \gev$, with a secondary minimum in the light Higgs
funnel region at $\mgl \sim 400 \gev$ and large $m_0$~\cite{mc4} that lies
beyond the present reach of the LHC~%
\footnote{A vestige of this region is visible in the VCMSSM at $\Delta \chi^2 \sim 7.5$, see
also the lower left panel of Fig.~\ref{fig:tBm12}.}.

\begin{figure*}[htb!]
\resizebox{8cm}{!}{\includegraphics{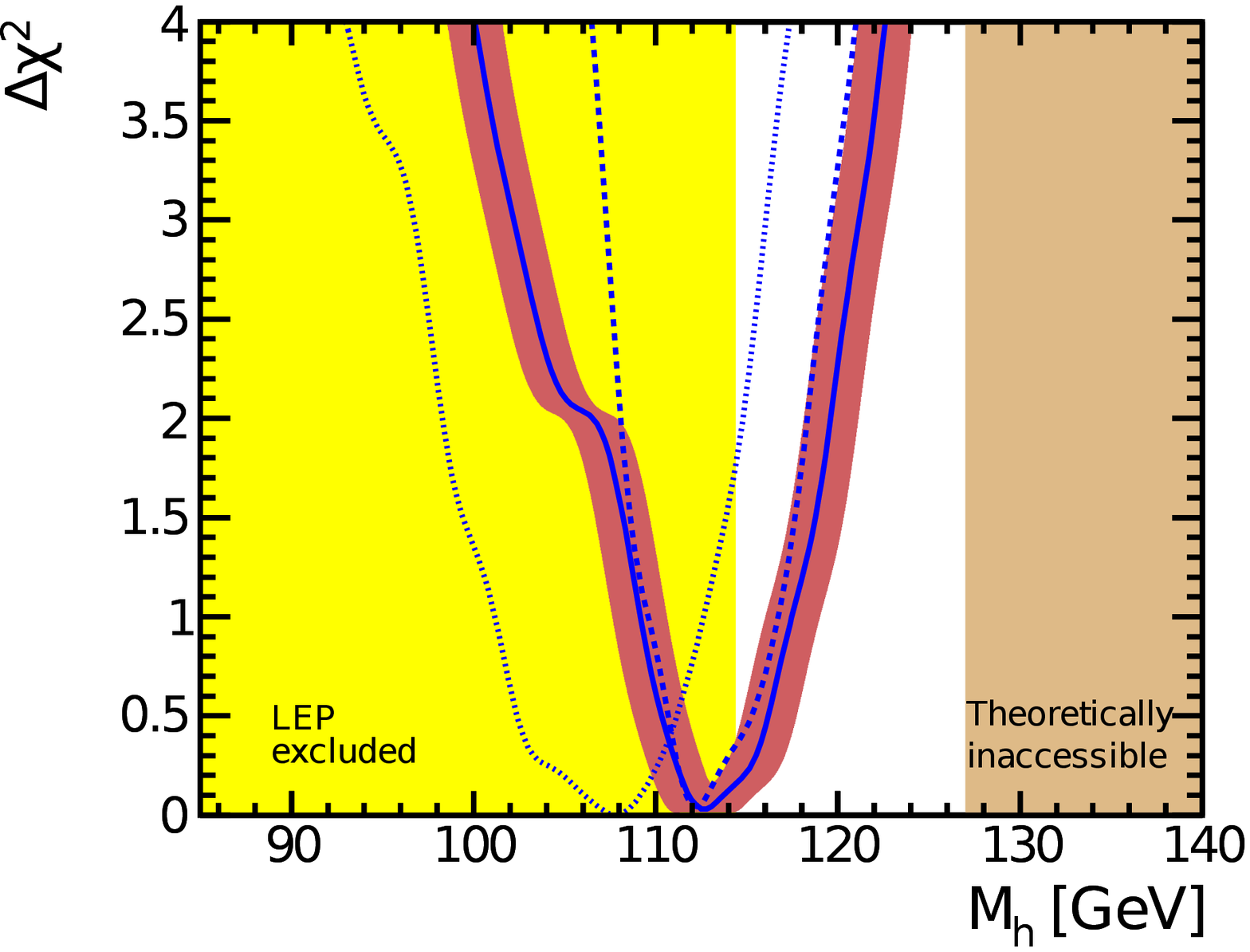}}
\resizebox{8cm}{!}{\includegraphics{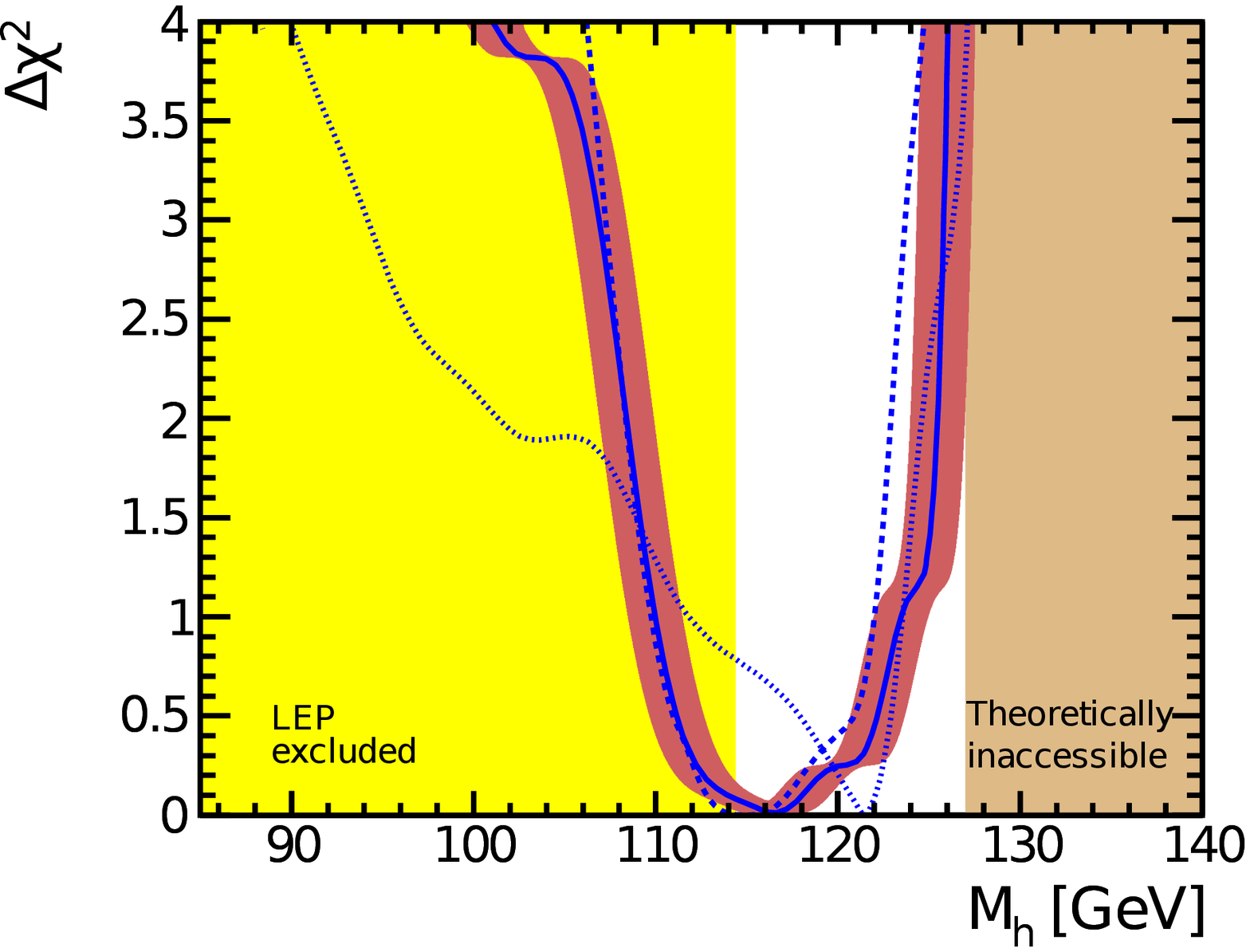}}
\resizebox{8cm}{!}{\includegraphics{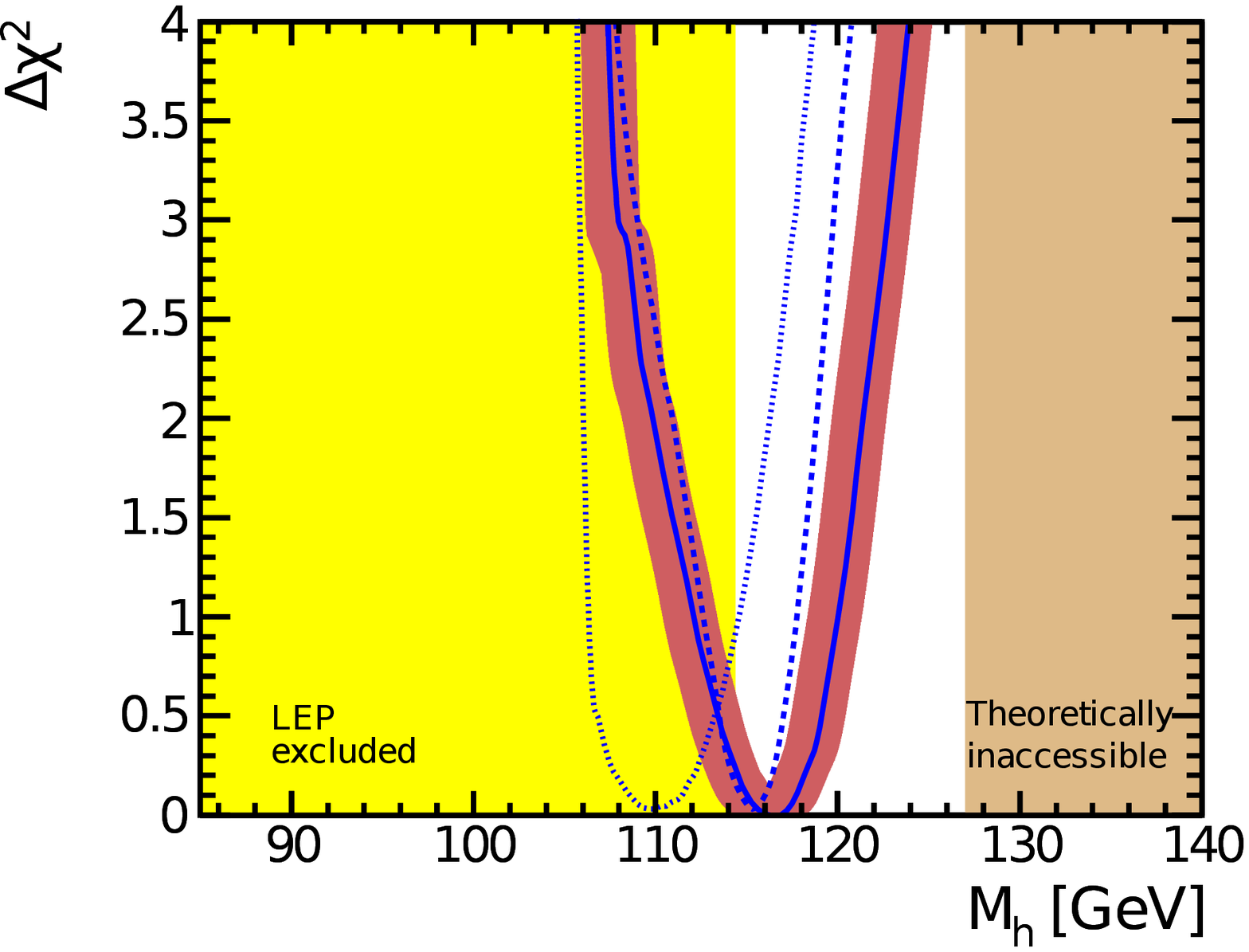}}
\resizebox{8cm}{!}{\includegraphics{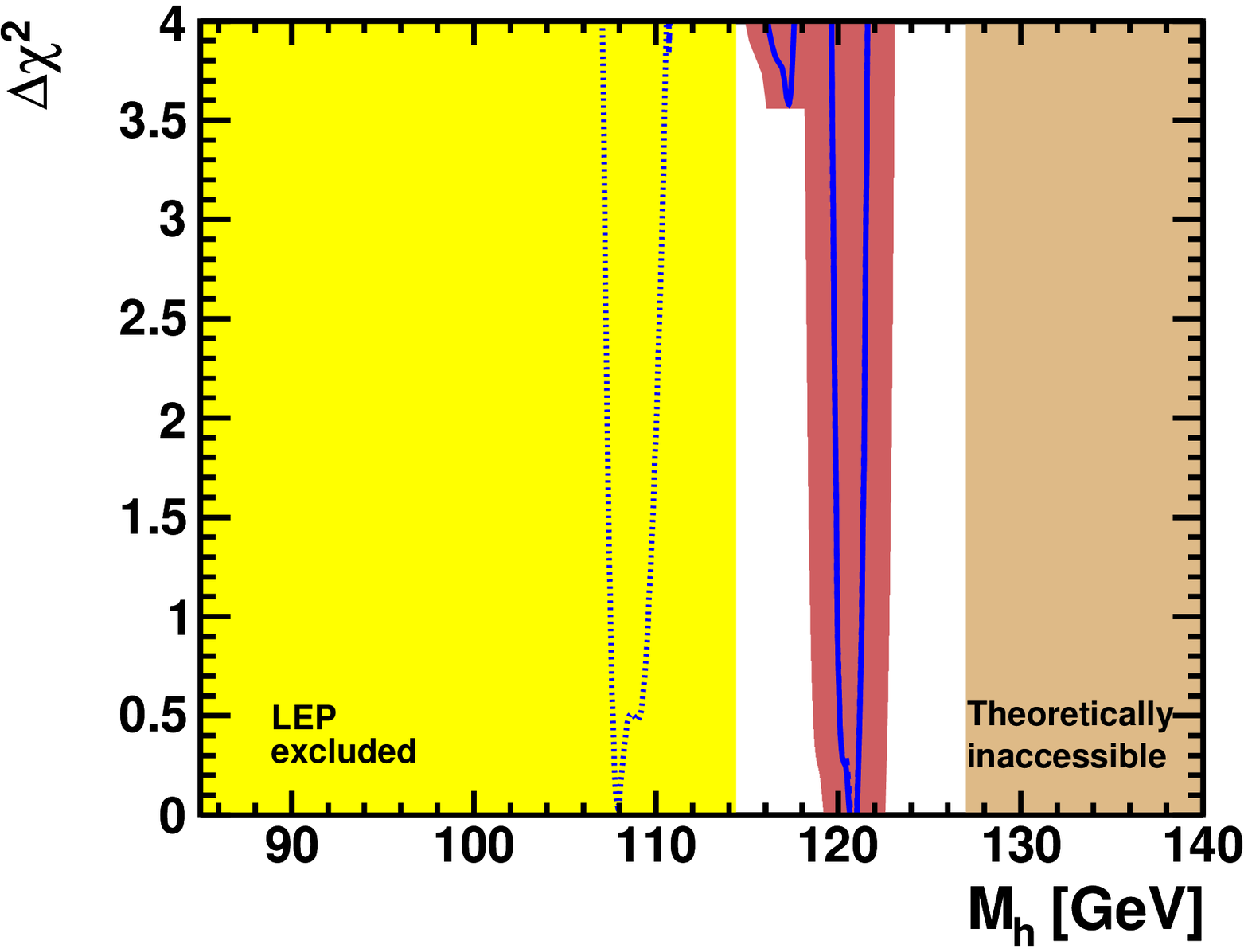}}\\
\caption{\it The one-parameter $\chi^2$ likelihood functions for the
  lightest MSSM Higgs mass $\Mh$ in the CMSSM (upper left),
NUHM1 (top right), VCMSSM (lower left) and mSUGRA (lower right). In each panel, we show
the $\chi^2$ functions including the \cmsatlas~\protect\cite{CMSsusy,ATLASsusy}
constraints as dashed (solid) lines, the latter with a red band indicating the estimated theoretical uncertainty 
in the calculation of $\Mh$
of $\sim 1.5 \gev$, and the pre-LHC $\chi^2$ function is shown as a dotted line.}
\label{fig:mh}
\end{figure*}

The experimental search for a SM-like Higgs boson is heating up, 
with interesting prospects for both the Tevatron collider~\cite{TeVH}
and the LHC~\cite{ATLAS,CMS}, see in particular~\cite{ATLASHiggs}. 
Accordingly, we display in
Fig.~\ref{fig:mh} the one-parameter $\chi^2$ functions for 
the lightest MSSM Higgs mass $\Mh$ in the CMSSM, NUHM1, VCMSSM 
and mSUGRA. In this figure we {\it do not\/} include the direct limits from
LEP~\cite{Barate:2003sz,Schael:2006cr} or the Tevatron, so as to 
illustrate whether there is a conflict between these 
limits and the predictions of supersymmetric models.
For each model we display the new likelihood functions incorporating
the ATLAS data as solid lines, indicating the
theoretical uncertainty in the calculation of $\Mh$ of $\sim 1.5 \gev$
by red bands. We also show, as dashed lines without red bands, the
central value of the prediction 
based on the CMS constraint, and as dotted lines without red bands the pre-CMS
predictions for $\Mh$ (all discarding the LEP constraint).

In the case of the CMSSM, we see that the CMS and
ATLAS constraints increase the consistency of the 
model prediction with the direct LEP limit on $\Mh$, indicated by the
yellow region: the best-fit value is found at $112.6 \; (112.8) \gev$
after the inclusion of the CMS (ATLAS) constraints (but still neglecting
the LEP Higgs searches, see also Table~\ref{tab:compare}), 
with an estimated theoretical error of $1.5 \gev$. 
In the case of the NUHM1, 
apart from somewhat lower best-fit values of 
$\Mh = 113.5 \; (116.5) \gev$ after including the
CMS (ATLAS) constraint (we recall that the $\chi^2$ function was
very shallow in the direction of lower masses),
we see that the main effect of the LHC data is
to increase substantially the one-parameter $\chi^2$ function at low masses  
$\Mh < 110 \gev$. (It should be remembered that in the NUHM1 the LEP
constraint is weakened at low $\Mh$ because the $hZZ$ coupling may be
reduced, which is not possible in the CMSSSM, VCMSSM and
  mSUGRA~\cite{Ellis:2001qv,Ambrosanio:2001xb}. The
LHC data render a large reduction less likely.) Now most of the preferred
$\Mh$ region in the NUHM1 is indeed above $\sim 114 \gev$.
In the case of the VCMSSM, the LHC constraint strongly disfavours a
Higgs boson below the LEP limit.
In the case of the mSUGRA coannihilation region,
the global minimum of the pre-LHC $\chi^2$ with the LEP constraint
disregarded was found in an isolated region at low $(m_0, m_{1/2})$, 
resulting in $\Mh \sim 108 \gev$. The inclusion of either the CMS or ATLAS data
removes this isolated region, pushing the preferred range to larger $\Mh \sim 121 \gev$
compatible with the LEP
constraint~\cite{Barate:2003sz,Schael:2006cr}. The global minimum with either
the CMS or ATLAS constraint included is very similar to the pre-CMS fit with
the LEP constraint included. In general, these plots
support the remark made earlier that the LHC data have an effect
comparable to the LEP $\Mh$ limit in constraining the models
studied. 

For this reason, the LHC data do not have a great impact on the amounts of
fine-tuning of parameters required. We have evaluated the fine-tuning measure proposed in~\cite{EENZ}
for the best pre- (post-)LHC fits, finding in the CMSSM 100 (120) [140] before the LHC (with the CMS constraint)
[with the ATLAS constraint]. The corresponding numbers in the other models are: NUHM 250 (230) [310], VCMSSM 130 (110) [140], mSUGRA 250 in all cases.~%
\footnote{See~\cite{Strumia} for a contrasting view that does not
take into account the LEP constraint on $\Mh$.}%

\begin{figure*}[htb!]
\resizebox{8cm}{!}{\includegraphics{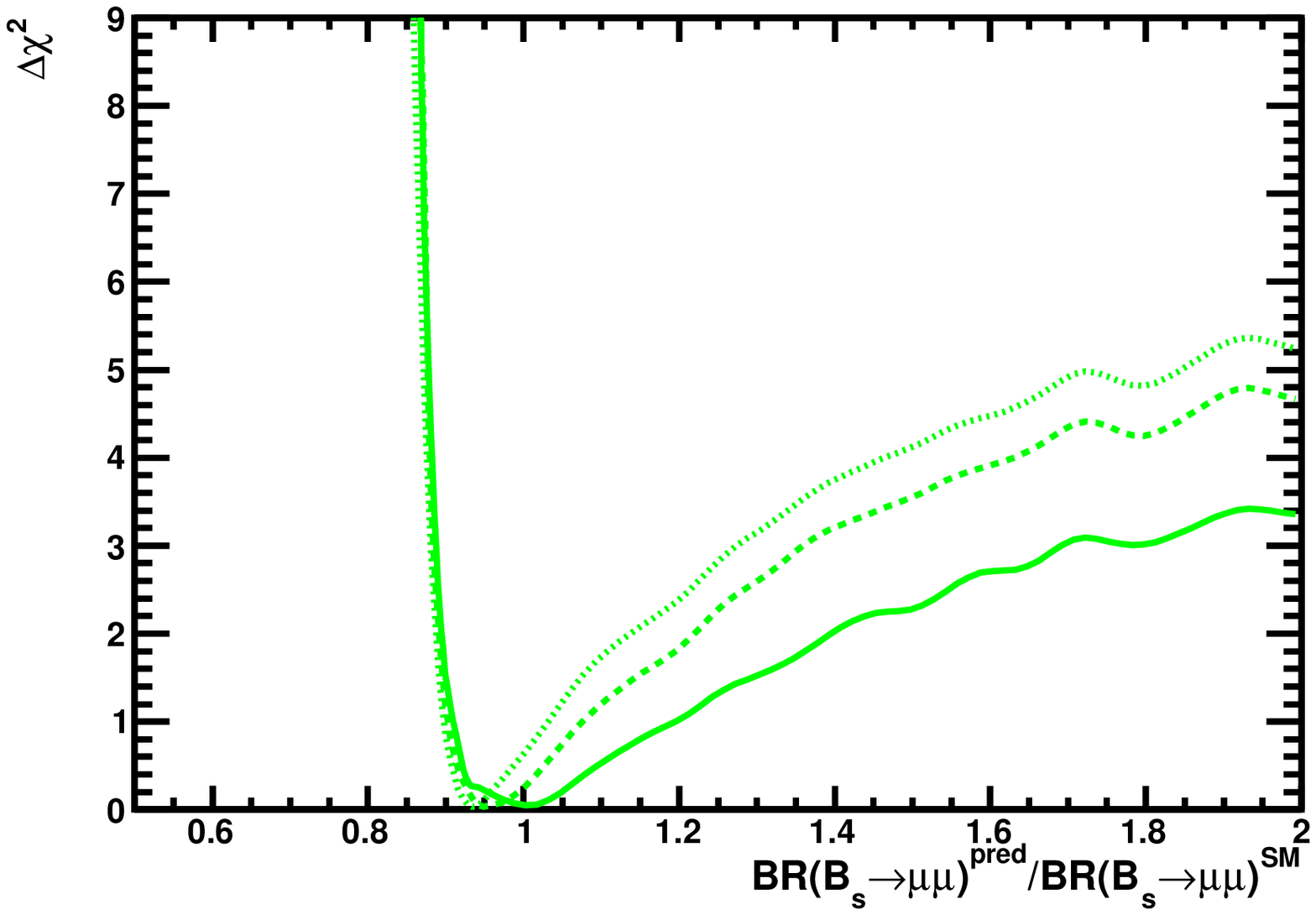}}
\resizebox{8cm}{!}{\includegraphics{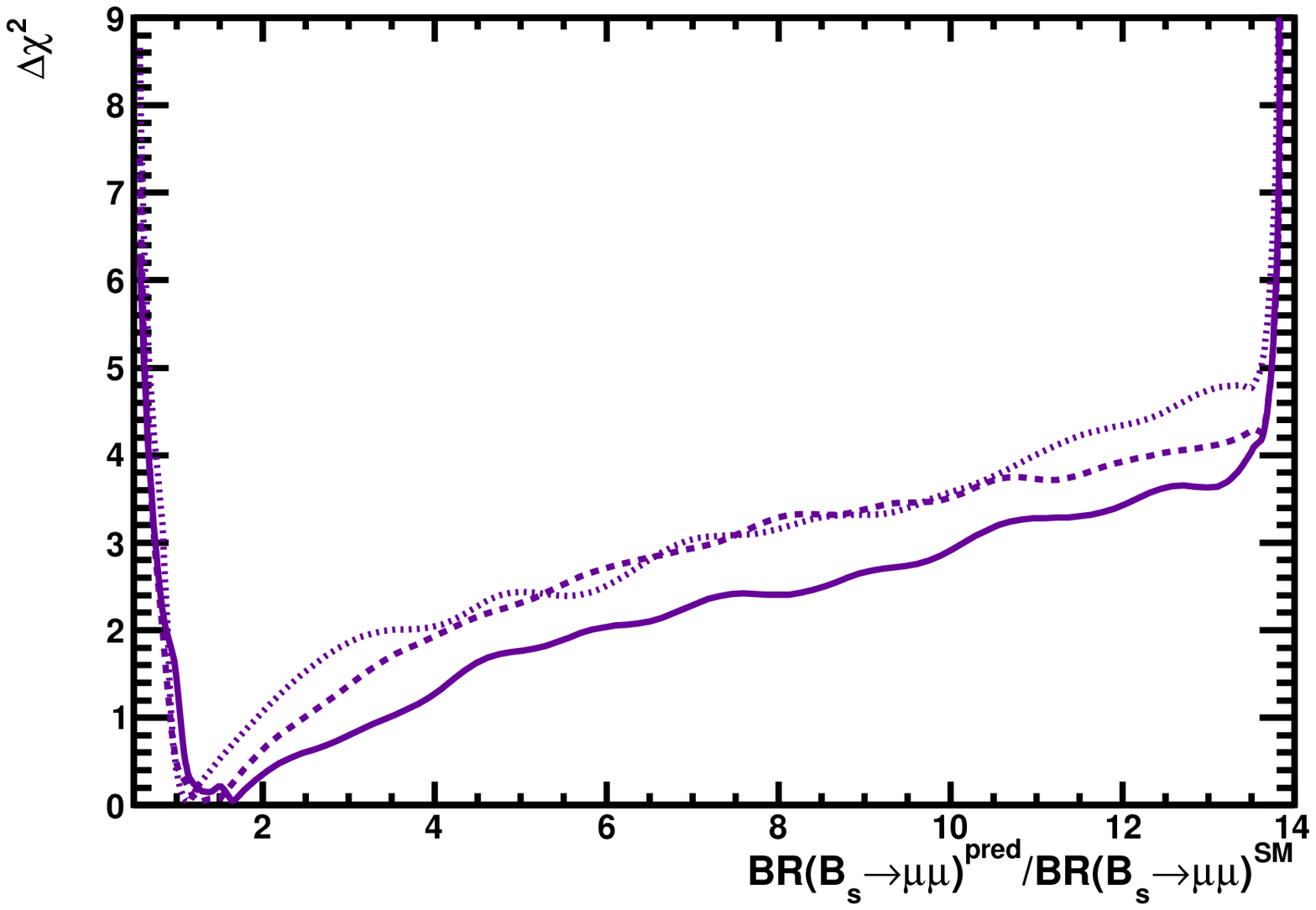}}
\resizebox{8cm}{!}{\includegraphics{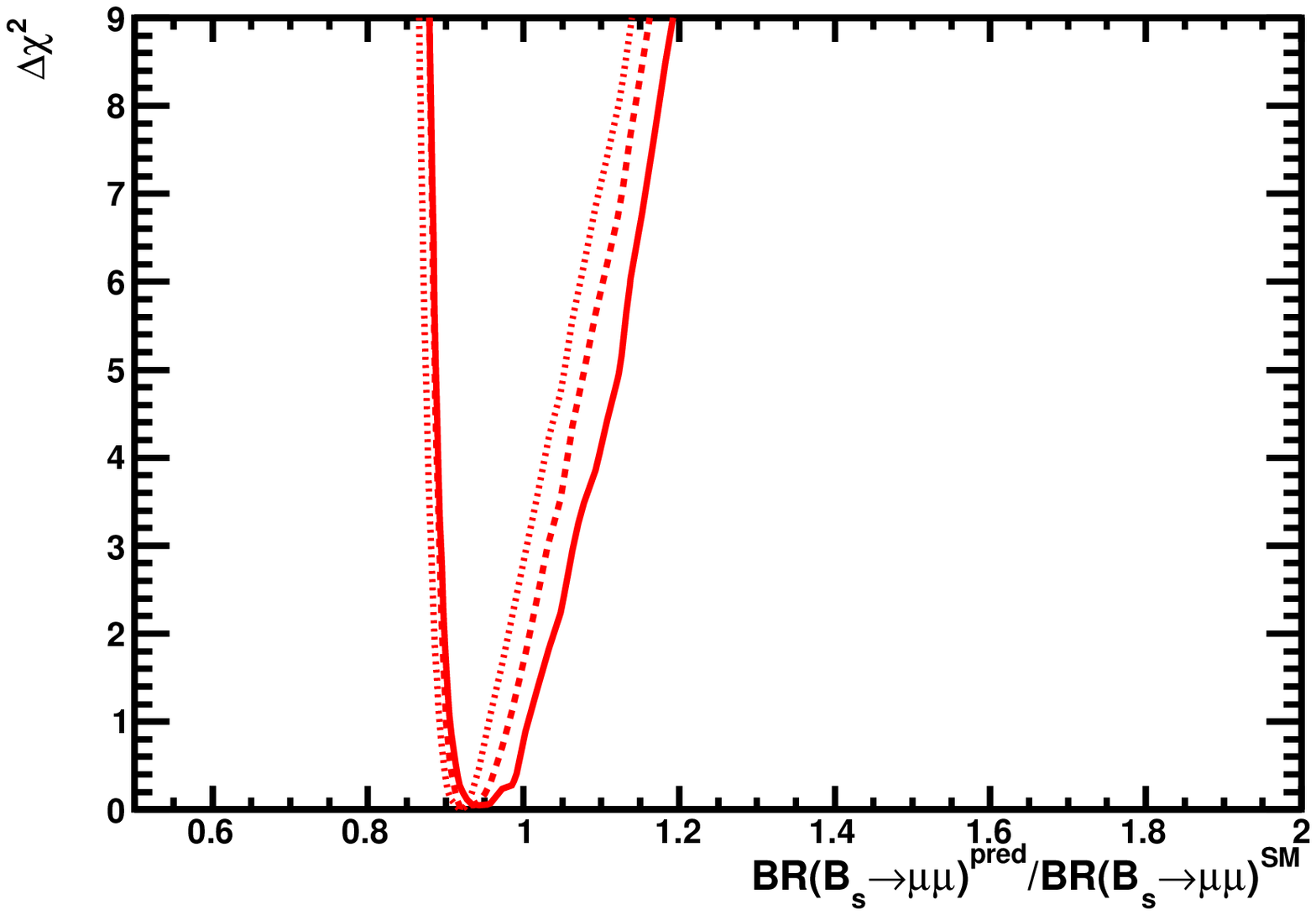}}
\resizebox{8cm}{!}{\includegraphics{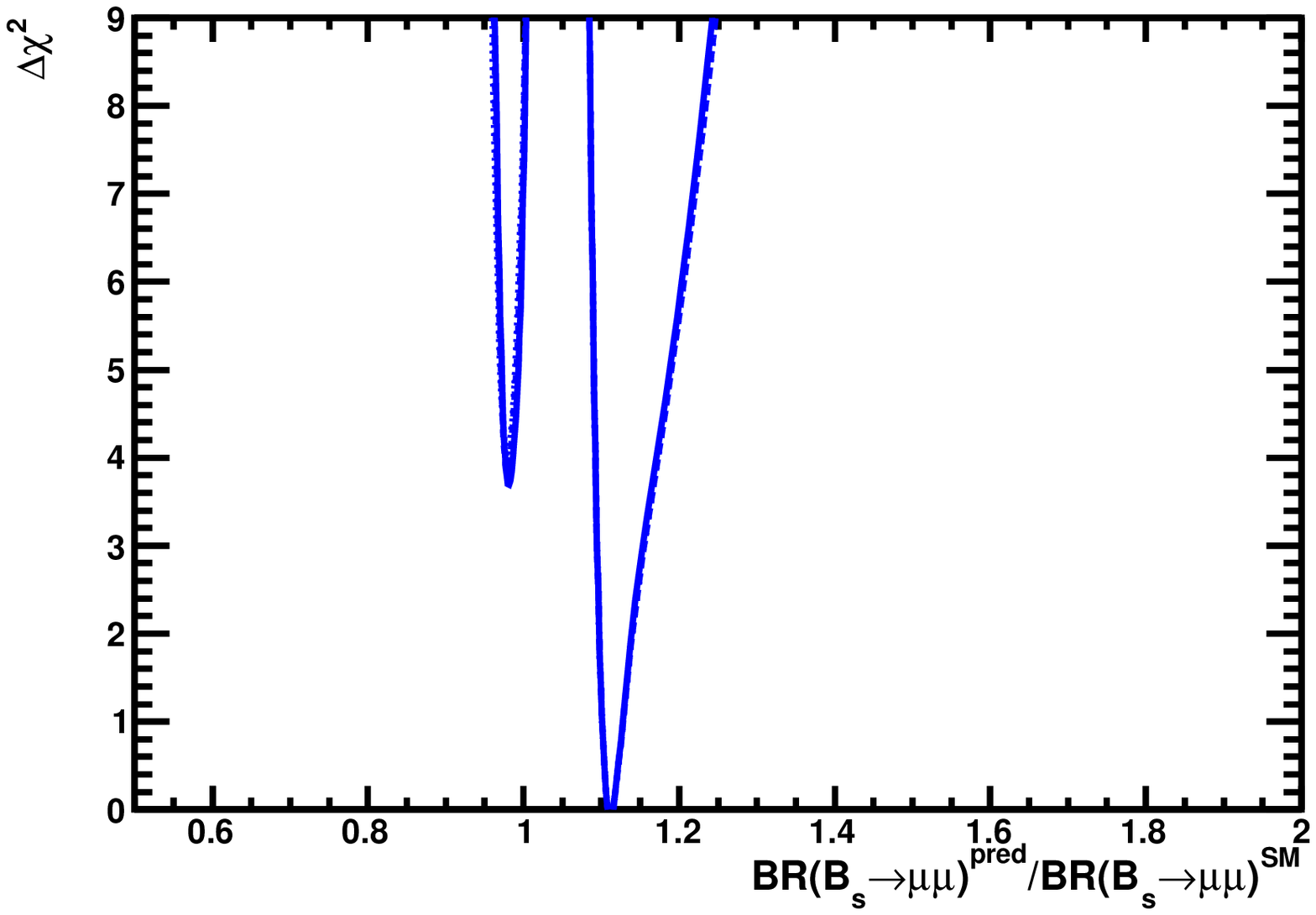}}\\
\vspace{-1cm}
\caption{\it The one-parameter $\chi^2$ likelihood functions for \bmm\ 
in the CMSSM (upper left),
NUHM1 (top right), VCMSSM (lower left) and mSUGRA (lower right).
In each panel, we show  the $\chi^2$ function including the
\cmsatlas~\protect\cite{CMSsusy,ATLASsusy} 
constraints as dashed and solid lines, respectively, and the previous
$\chi^2$ function as a dotted line.}
\label{fig:bsmumu}
\end{figure*}

One of the other observables of potential interest for testing
variants of the MSSM is \bmm, and we recall that the sensitivity
of the LHCb experiment from the data recorded during the 2010
LHC run approaches that already achieved by the CDF and D\O\
experiments~\cite{LHCbbmm}. As seen in Fig.~\ref{fig:bsmumu},
we find that the effects of the LHC data on this
observable are negligible in mSUGRA and very small in the VCMSSM. However,
the larger values of $\tan \beta$ now favoured in the CMSSM increase the likelihood of an enhancement
in \bmm\ beyond the SM value.
As was discussed previously~\cite{mc4}, the best prospects for a deviation from the SM
prediction for \bmm\ are in the NUHM1, where the best-fit value is significantly
larger than in the SM and substantially larger values are now
possible, thanks to the larger values of $\tan \beta$ now preferred.
On the other hand, the best-fit values for \bmm\
in the CMSSM, VCMSSM and mSUGRA are peaked within
10\% of the SM value, indistinguishable from the SM given the
theoretical and experimental uncertainties.

\begin{figure*}[htb!]
\resizebox{8cm}{!}{\includegraphics{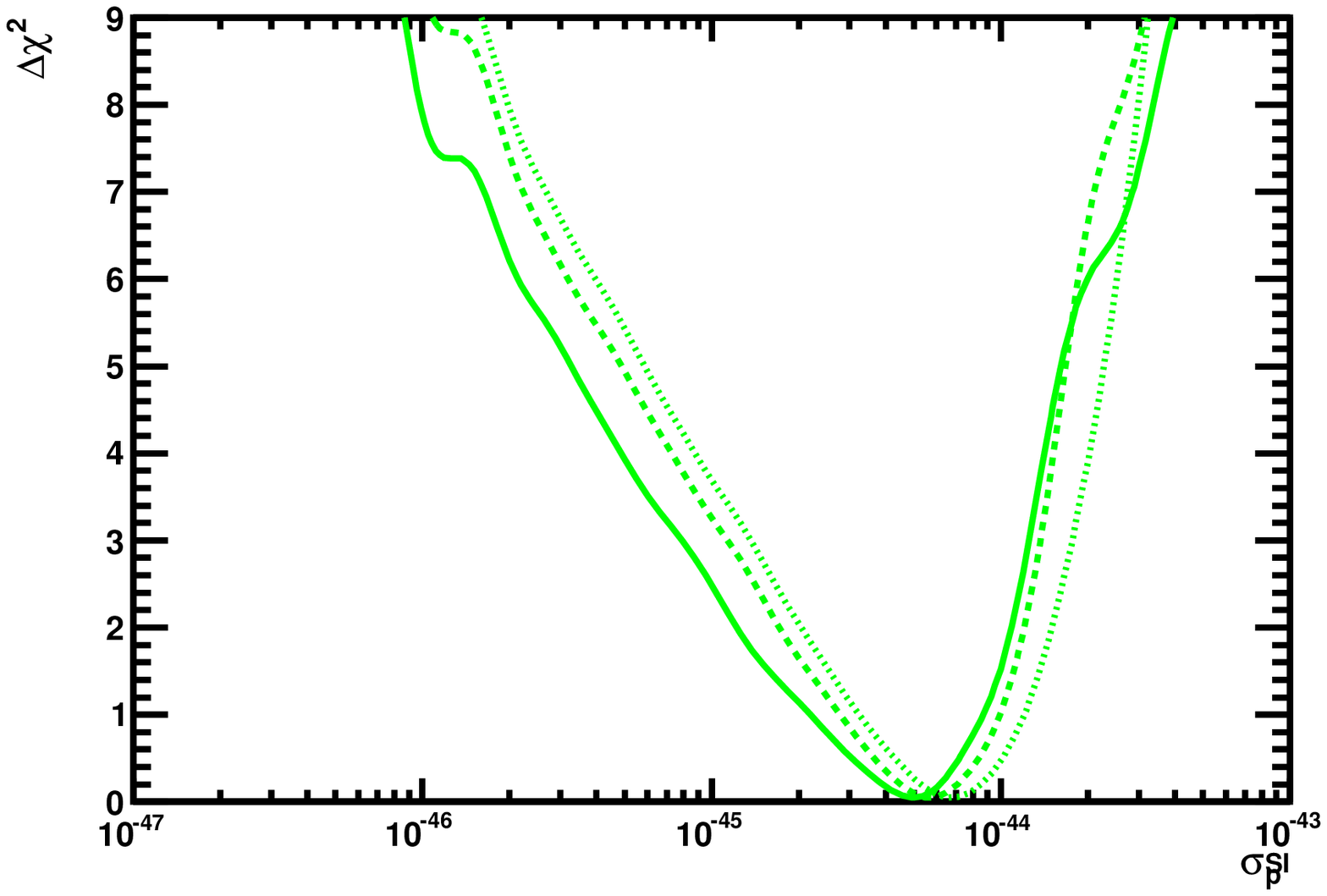}}
\resizebox{8cm}{!}{\includegraphics{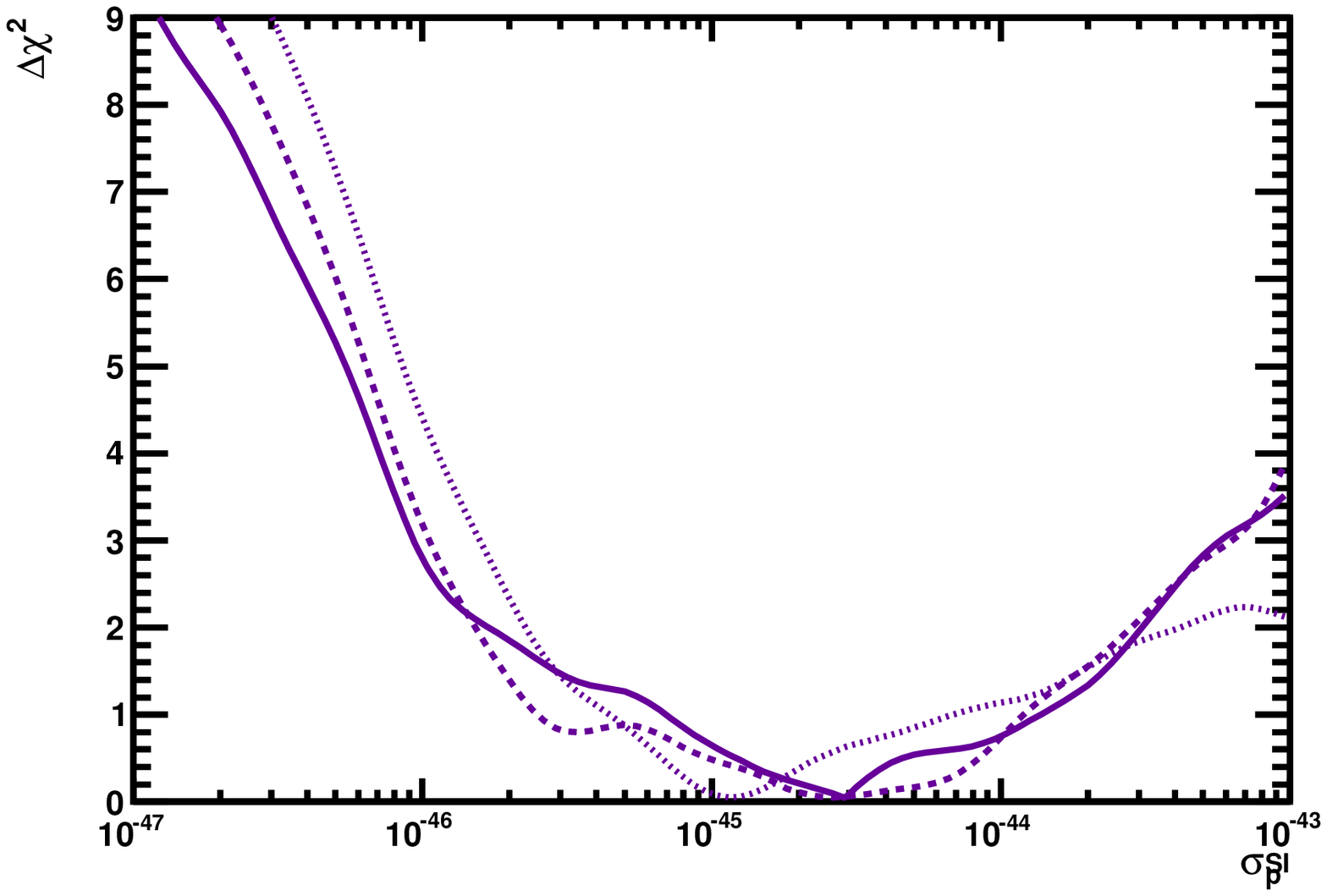}}
\resizebox{8cm}{!}{\includegraphics{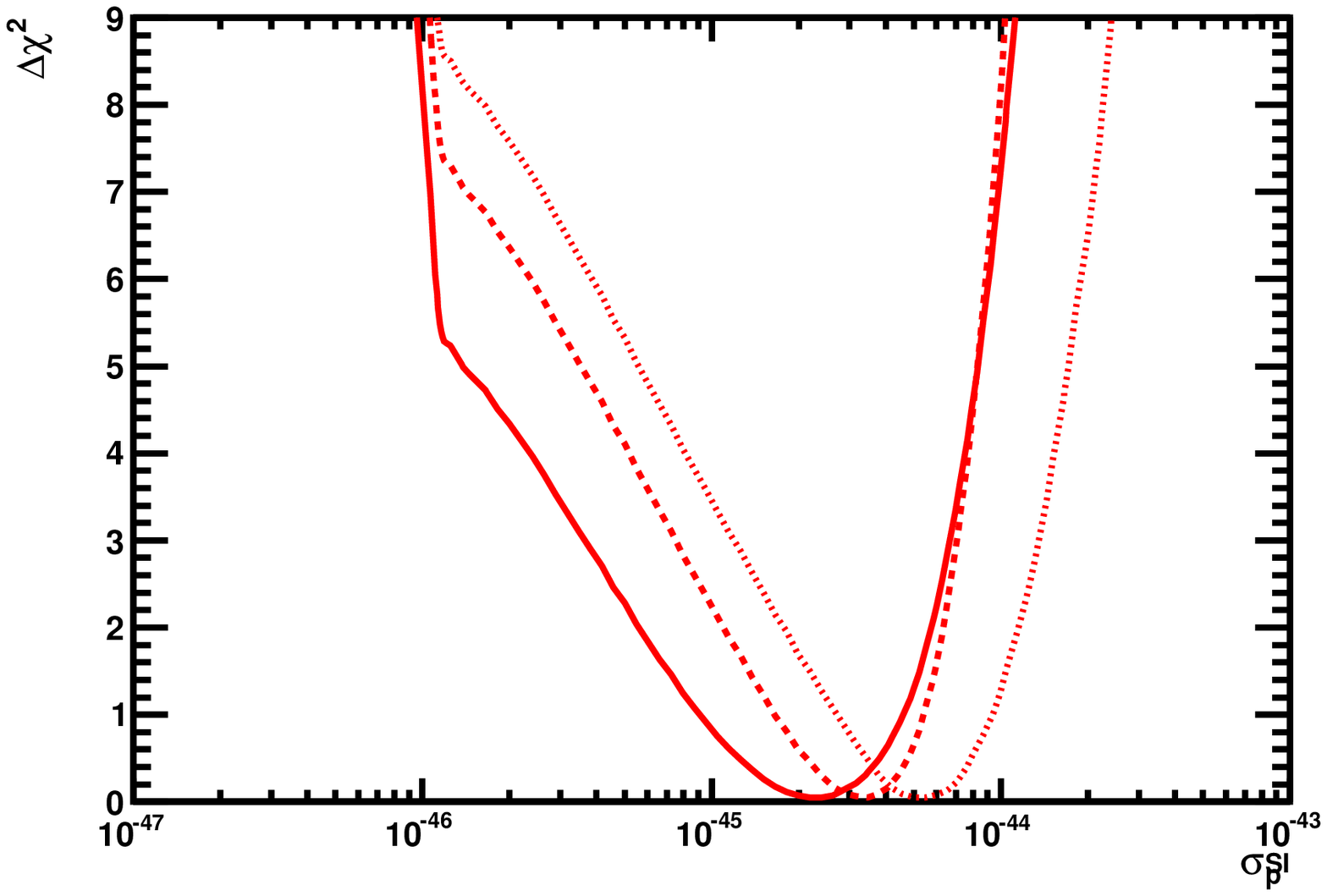}}
\resizebox{8cm}{!}{\includegraphics{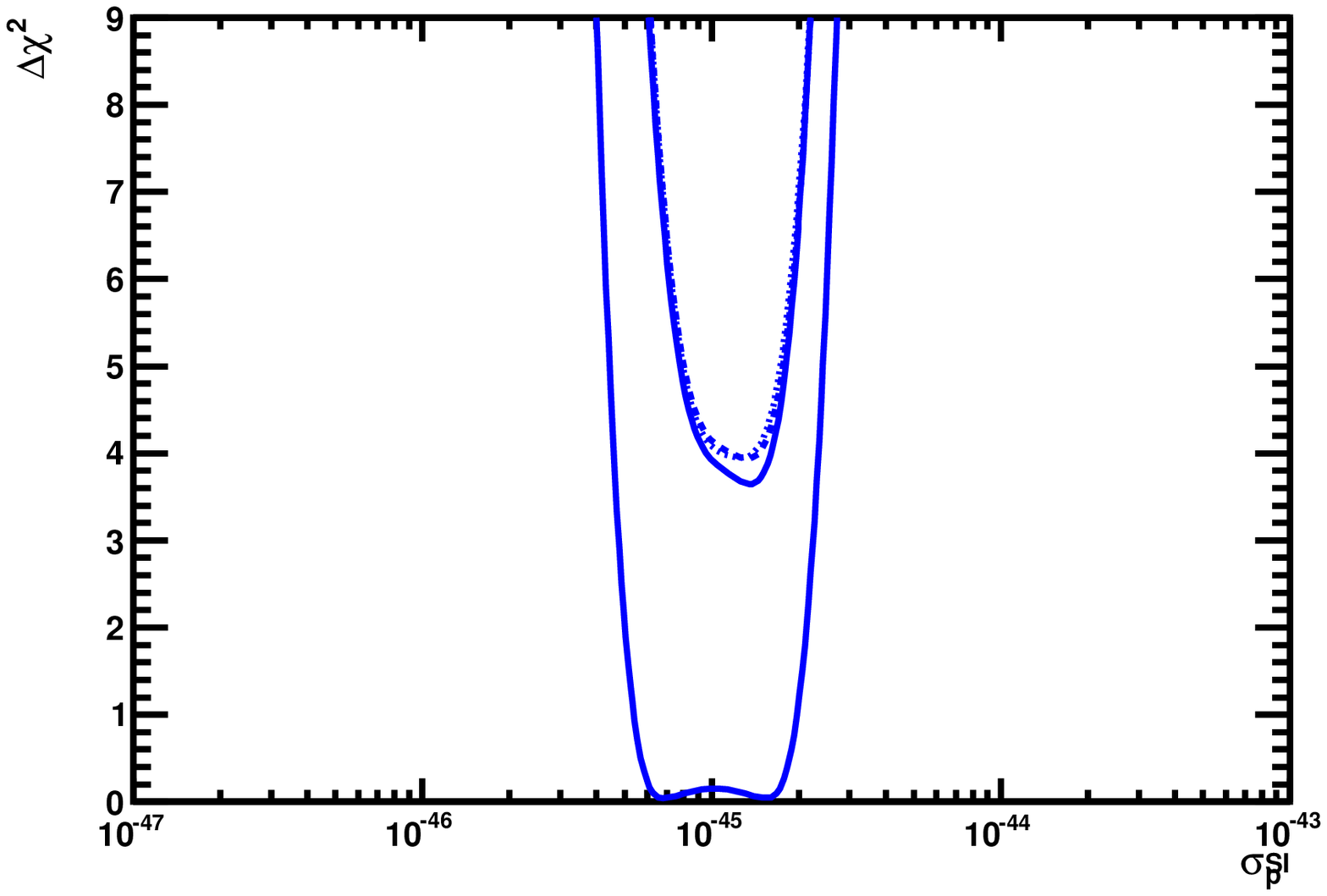}}\\
\vspace{-1cm}
\caption{\it The one-parameter $\chi^2$ likelihood functions for the
  spin-independent neutralino dark matter scattering cross section
  $\ssi$ in the CMSSM (upper left), NUHM1 (top right), VCMSSM (lower
  left) and mSUGRA (lower right). In each panel, we show  the $\chi^2$
  function calculated assuming
$\Sigma_{\pi N} = 64 \mev$ and including the
  \cmsatlas~\protect\cite{CMSsusy,ATLASsusy} constraints 
as dashed and solid lines, and the previous $\chi^2$ function for
  $\Sigma_{\pi N}$ as a  dotted line.
}  
\label{fig:ssi}
\end{figure*}

Finally, we recall that it is expected that the sensitivities of direct
searches for the spin-independent scattering of dark matter particles on
heavy nuclei will soon be increasing
substantially~\cite{Xenon100}. Accordingly, we display in
Fig.~\ref{fig:ssi} the one-parameter $\chi^2$ functions for the
spin-independent neutralino dark matter scattering cross section $\ssi$
in the CMSSM, NUHM1, VCMSSM  and mSUGRA.
We display the
new likelihood functions incorporating \cmsatlas\ 
data and calculated assuming
the default value $\Sigma_{\pi N} = 64$~MeV~\cite{EOSavage}
as dashed and solid lines, as previously, and those given by the
previous fits as dotted lines.
We see that in the CMSSM and VCMSSM the 
LHC constraint reduces substantially the likelihood of relatively
large values of $\ssi > 10^{-44}$~cm$^2$ for $\Sigma_{\pi N} = 64 \mev$, whereas the likelihood
functions are little changed in mSUGRA and the NUHM1, due in the latter case to compensation
between preferences for larger $m_{1/2}$ and larger $\tan \beta$.
We have recalculated the values of $\ssi$ at the best-fit points in the
different models assuming $\Sigma_{\pi N} = 45 \mev$, close to the lower
end of the plausible range~\cite{EOSavage},  
and found reductions in $\ssi$ by factors $\sim 3$ to 4 in each case.

\begin{figure*}[htb!]
\resizebox{8cm}{!}{\includegraphics{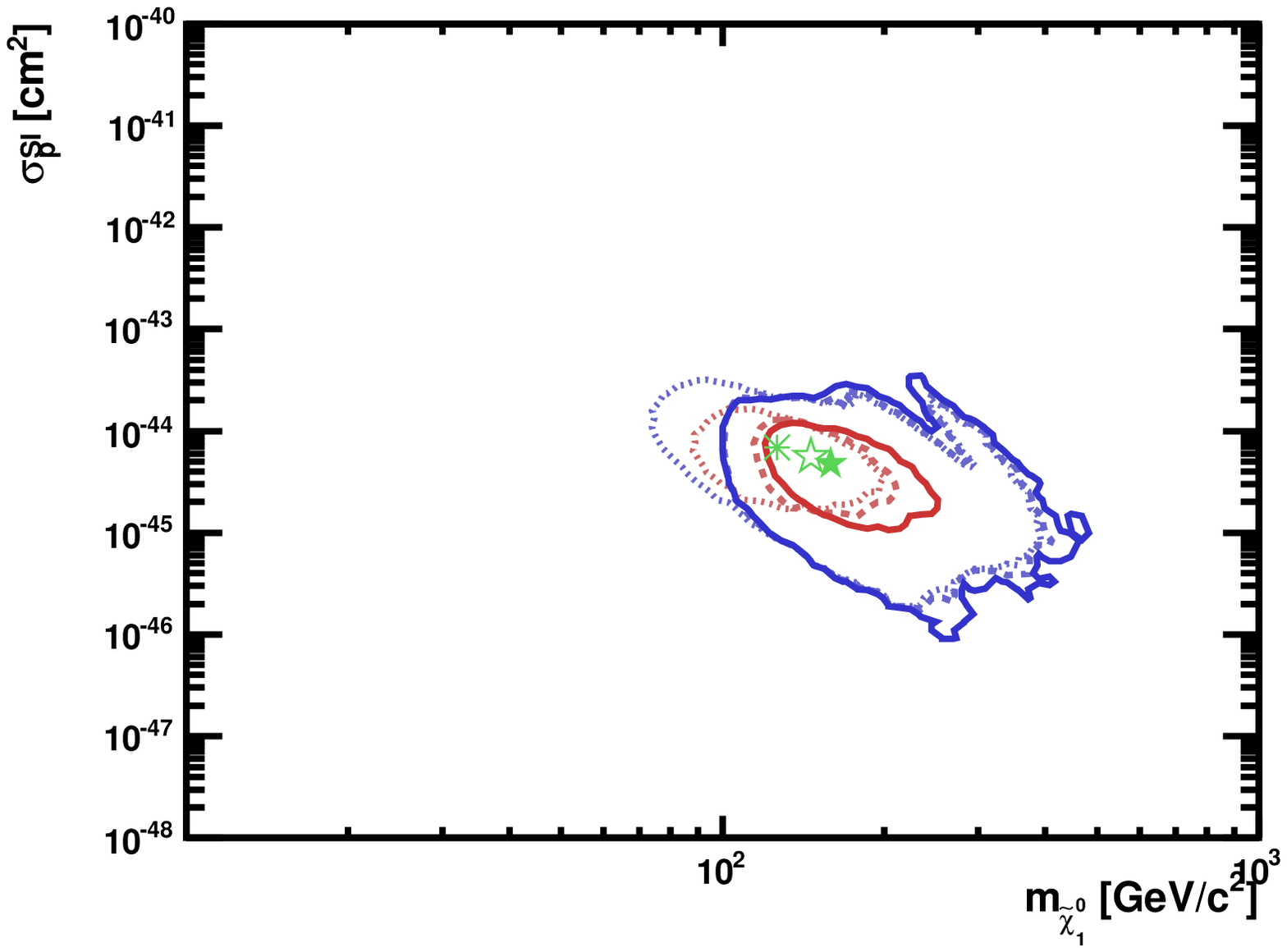}}
\resizebox{8cm}{!}{\includegraphics{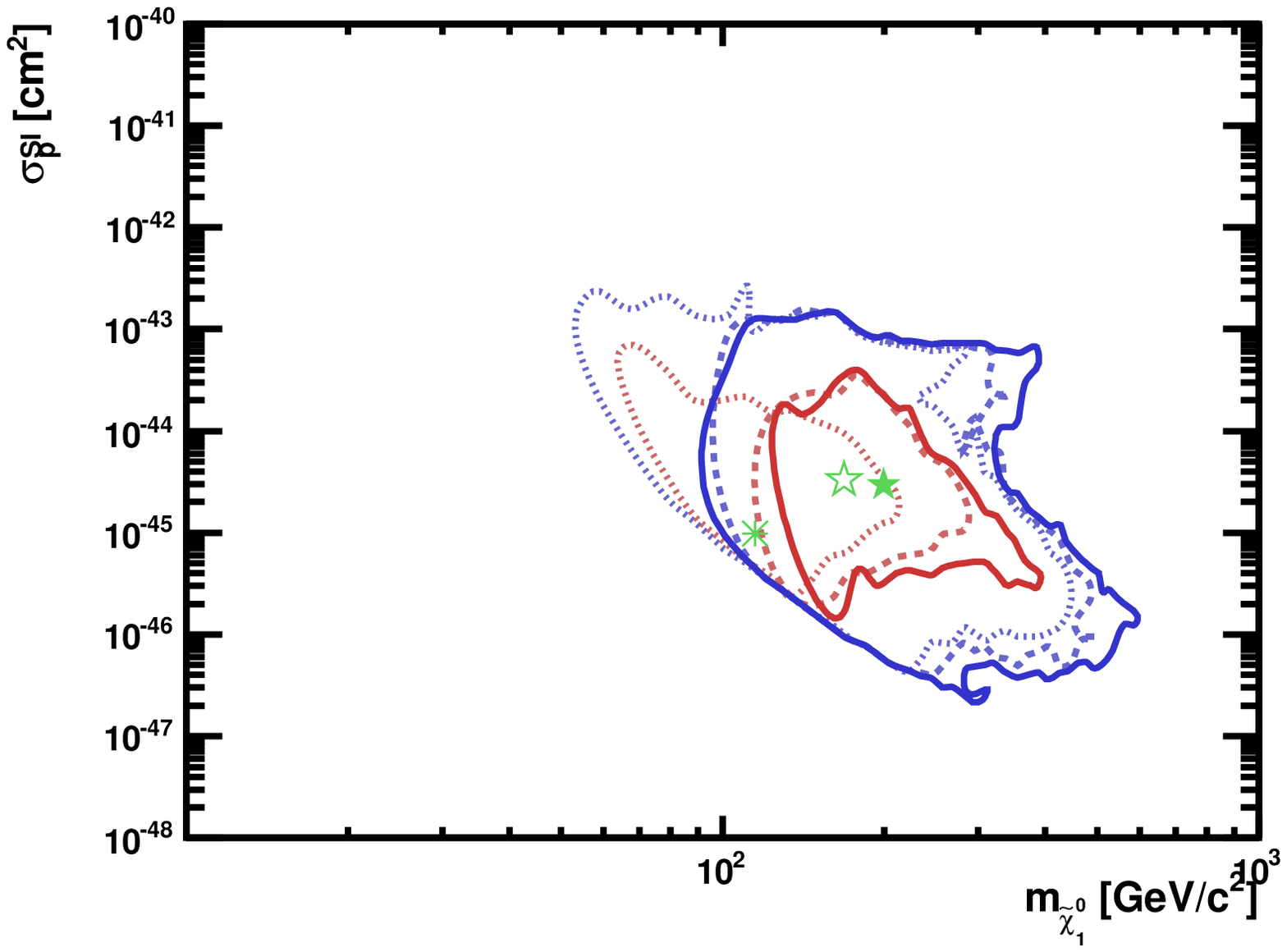}}
\resizebox{8cm}{!}{\includegraphics{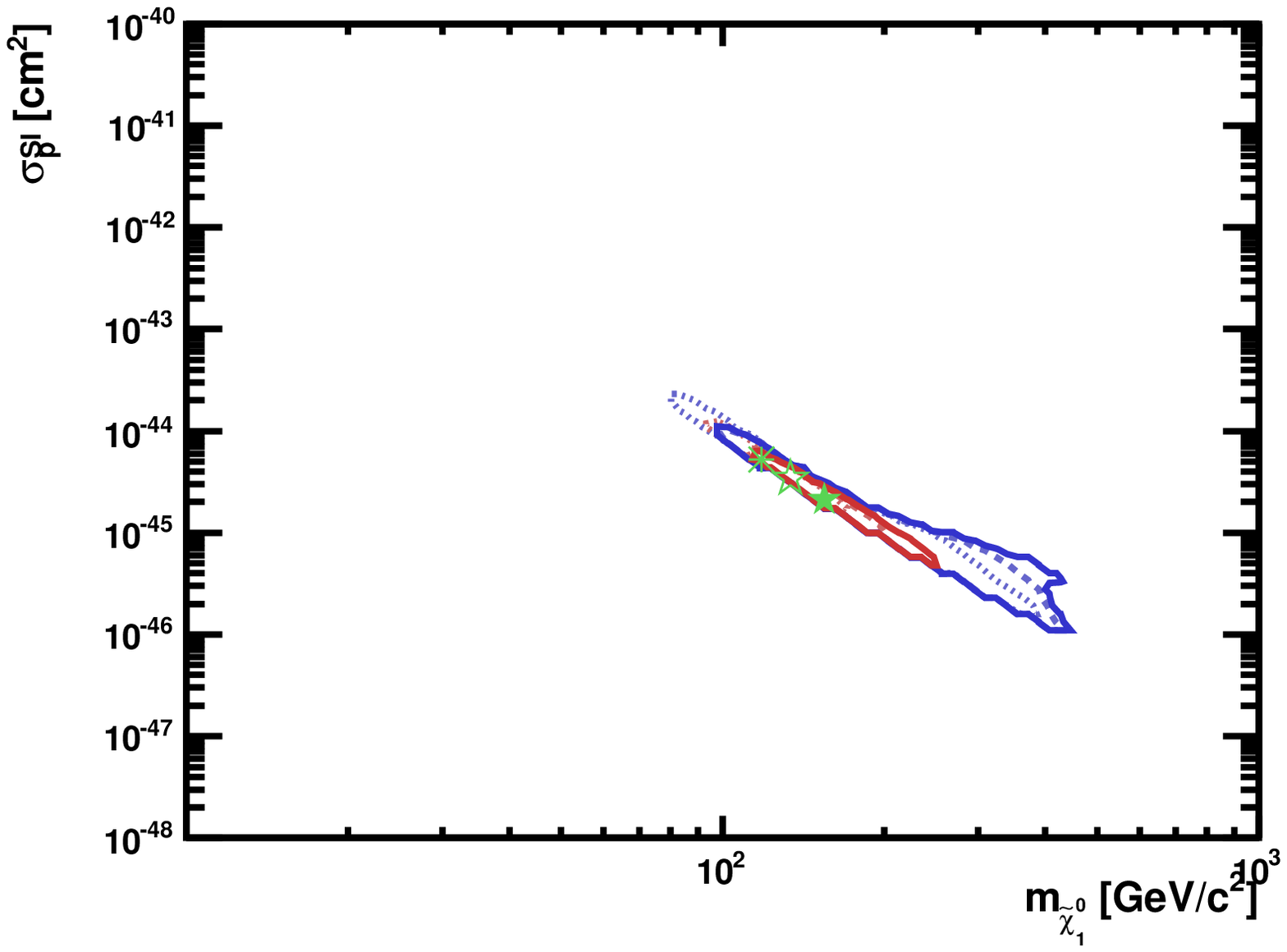}}
\resizebox{8cm}{!}{\includegraphics{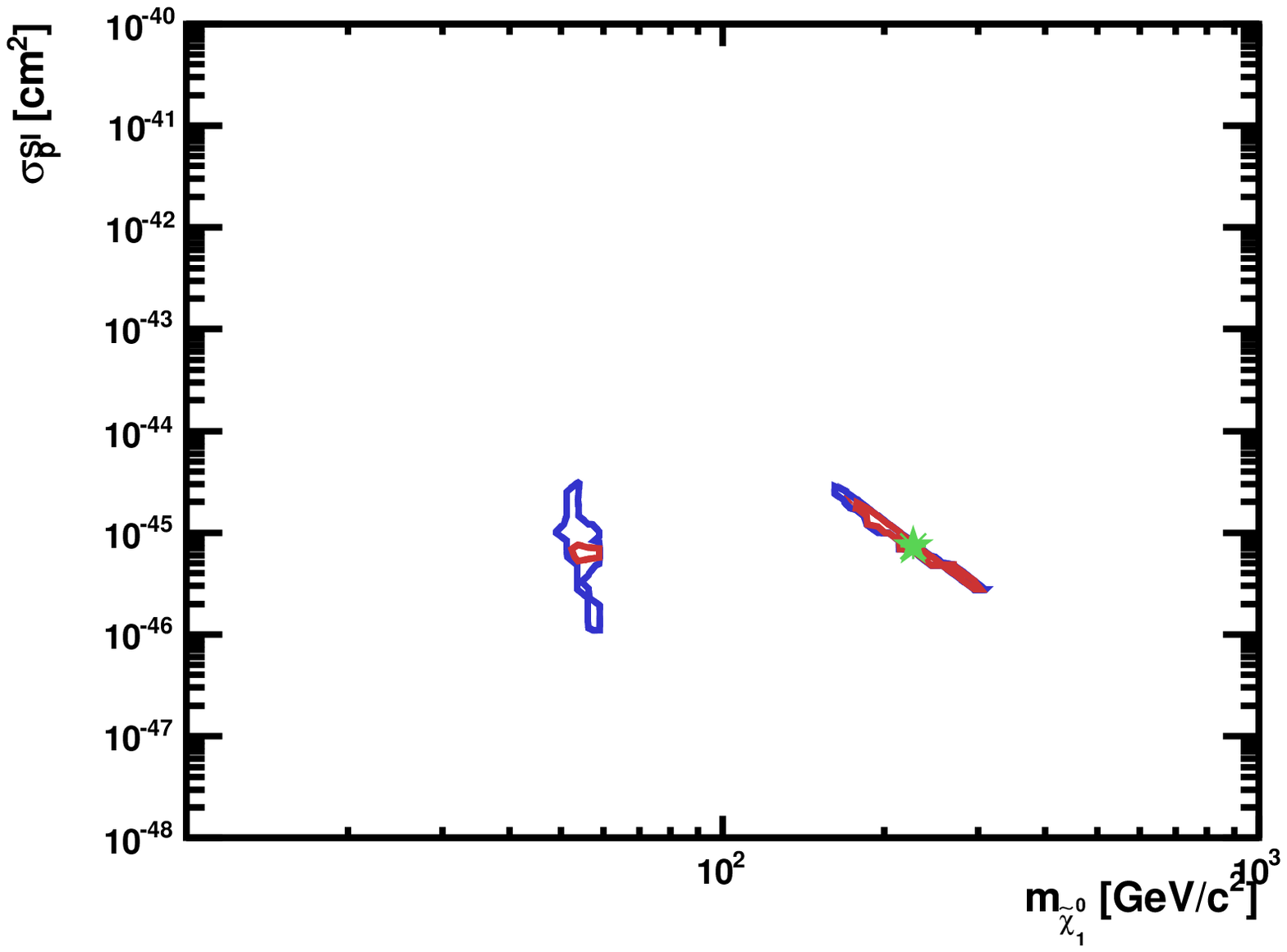}}\\
\vspace{-1cm}
\caption{\it The regions of the $(\mneu{1}, \ssi)$ planes in the 
CMSSM (upper left), NUHM1 (top right), VCMSSM (lower
left) and mSUGRA (lower right) favoured at
the 68\% and 95\% CL including the CMS constraint (solid lines)
and excluding it (dashed lines). The best-fit points after (before)
the CMS and ATLAS constraints~\protect\cite{CMSsusy,ATLASsusy} are shown as open (solid) green stars,
and the best pre-LHC fits as green `snowflakes'.
The results are calculated assuming
$\Sigma_{\pi N} = 64 \mev$.
} 
\label{fig:ssimneu}
\end{figure*}

Fig.~\ref{fig:ssimneu} displays the regions of the $(\mneu{1}, \ssi)$ planes
favoured at the 68\% and 95\% CL in the CMSSM, NUHM1, VCMSSM and mSUGRA
models, assuming $\Sigma_{\pi N} = 64 \mev$, comparing the post- and
pre-LHC contours (dashed/solid lines and dotted lines, respectively),
and showing the best-fit points as solid and open stars, 
respectively~%
\footnote{The region at small $\mneu{1}$ in the mSUGRA plot
corresponds to the light Higgs funnel discussed in~\cite{mc4}.}%
. The models predict, in general, a negative correlation between
$\mneu{1}$ and $\ssi$, which is why the stronger lower limit on $\mgl$
and hence $\mneu{1}$ provided by CMS and ATLAS (see Fig.~\ref{fig:mgl})
corresponds to a stronger upper limit on $\ssi$, as seen in
Fig.~\ref{fig:ssi}. Conversely, any future upper limit on (measurement
of) $\ssi$ would correspond to a lower limit on (preferred range of) 
$\mneu{1}$ and hence $\mgl$. However, this correlation is weakened by the
uncertainty in $\Sigma_{\pi N}$, which should be taken into account in
the interpretation of any future constraint on $\ssi$.

\bigskip
In summary, we have shown that the initial \cmsatlas\ searches for
supersymmetry at the LHC at 7~TeV already have significant impacts on
the regions of the 
CMSSM, NUHM1 and VCMSSM parameter spaces favoured at the 68 and 95\%~CL,
shifting the best-fit points to somewhat larger values (with an
estimated systematic uncertainty of $\sim 10\%$ (CMS) to 20\% (ATLAS) in $m_{1/2}$). They are
now located at 
$(m_{1/2}, m_0) \sim (340~{\rm to}~490, 100) \gev$, corresponding to $\mgl \sim 800~{\rm to}~1000 \gev$,
within the prospective reach of the LHC with 1/fb of analyzed data at a
centre-of-mass energy of 7~TeV or more~%
\footnote{However, we recall that significantly larger values
of $\mgl$ are allowed at the 95\%~CL.}. 
On the other hand, the \cmsatlas\ data have no impact on the mSUGRA fit.
Increases in the preferred values of $\tan \beta$ in the CMSSM and particularly the NUHM1
increase the likelihood that \bmm\ exceeds significantly the SM value.
The model predictions for $\Mh$ (discarding the direct searches by LEP)
are now in better agreement with the direct search limits, because of the
somewhat heavier mass spectrum required by CMS and ATLAS. 
Within the NUHM1, the new $\chi^2$ contribution decreases significantly
the likelihood of a light Higgs boson with $\Mh \lsim 110 \gev$. 
Within the CMSSM and VCMSSM frameworks, the \cmsatlas\ results also 
diminish significantly the likelihood that $\ssi > 10^{-44}$~cm$^2$, but have
little impact in the NUHM1 and mSUGRA.

The search for supersymmetry is about to enter its critical stage, and
further results from the LHC with tailored analyses and increased
luminosity will soon provide crucial new information, 
as will dark matter search experiments.


\bigskip
\subsection*{Note added}

During the completion of this work
three papers addressing related topics appeared~\cite{Nath,Allanach,Bottino}. 
Ref.~\cite{Nath}
studies the phenomenology of the mSUGRA rapid-annihilation funnel region. Ref.~\cite{Allanach}
analyzes the CMS constraint within the CMSSM and reaches similar conclusions on the increases in
$m_{1/2}$ and $\tb$ that it entails. Ref.~\cite{Bottino} considers a more general model than those examined here.

\bigskip
This work was supported in part by the European Community's Marie-Curie
Research Training Network under contracts MRTN-CT-2006-035505
`Tools and Precision Calculations for Physics Discoveries at Colliders'
and MRTN-CT-2006-035482 `FLAVIAnet', and by the Spanish MEC and FEDER under 
grant FPA2005-01678. The work of S.H. was supported 
in part by CICYT (grant FPA~2007--66387 and FPA 2010--22163-C02-01), and
the work of K.A.O. was supported in part
by DOE grant DE--FG02--94ER--40823 at the University of Minnesota.
K.A.O. also thanks SLAC 
 (supported by the DOE under contract number DE-AC02-76SF00515) and 
 the Stanford Institute for Theoretical Physics
 for their hospitality and support while this work was being finished.



\begin{thebibliography}{99}


\bibitem{ATLAS}
G.~Aad {\it et al.} ATLAS Collaboration,
{\it Expected Performance of the ATLAS Experiment - Detector, Trigger and
  Physics,}
  arXiv:0901.0512.

\bibitem{CMS}
G.~L.~Bayatian {\it et al.}, CMS Collaboration,  
{\it CMS Technical Design Report, Volume II: Physics Performance}, 
CERN-LHCC-2006-021, CMS-TDR-008-2 
J.~Phys.~G34, 995 (2007); see: {\tt http://cmsdoc.cern.ch/cms/cpt/tdr/}~.

 \bibitem{EHNOS}
J. Ellis, J.S. Hagelin, D.V. Nanopoulos, K.A. Olive
and M. Srednicki, Nucl. Phys. B {\bf 238} (1984) 453; 
H. Goldberg, Phys. Rev. Lett. {\bf 50} (1983) 1419.

\bibitem{hierarchy}
L.~Maiani, {\it All You Need To Know About The Higgs Boson}, Proceedings
of the Gif-sur-Yvette Summer School On Particle Physics, 1979, pp.1-52; G.~'t~Hooft, in {\it
Recent developments in Gauge Theories}, Proceedings of the NATO Advanced Study
Institute, Carg{\`e}se, 1979, eds. G.~'t~Hooft et al. (Plenum Press, NY, 1980);
E.~Witten,
  Phys.\ Lett.\ B {\bf 105} (1981) 267.

\bibitem{newBNL} 
 The Muon g-2 Collaboration,
 {\it Phys. Rev. Lett.} {\bf 92} (2004) 161802.
 [arXiv:hep-ex/0401008];
 G.~Bennett et al.\ [The Muon g-2 Collaboration],
  {\em Phys.\ Rev.} {\bf D 73} (2006) 072003
  [arXiv:hep-ex/0602035].

\bibitem{g-2}
  D.~Stockinger,
  J.\ Phys.\ G {\bf 34} (2007) R45
  [arXiv:hep-ph/0609168];
    J.~Miller, E.~de~Rafael and B.~Roberts,
   {\em Rept.\ Prog.\ Phys.} {\bf 70} (2007) 795
   [arXiv:hep-ph/0703049];
    J.~Prades, E.~de Rafael and A.~Vainshtein,
  arXiv:0901.0306 [hep-ph];
    F.~Jegerlehner and A.~Nyffeler,
  Phys.\ Rept.\  {\bf 477}, 1 (2009)
  [arXiv:0902.3360 [hep-ph]];
  J.~Prades,
  Acta Phys.\ Polon.\ Supp.\  {\bf 3}, 75 (2010)
  [arXiv:0909.2546 [hep-ph]];
    T.~Teubner, K.~Hagiwara, R.~Liao, A.~D.~Martin and D.~Nomura,
  arXiv:1001.5401 [hep-ph].

  \bibitem{newDavier}
M.~Davier, A.~Hoecker, B.~Malaescu and Z.~Zhang,
  arXiv:1010.4180 [hep-ph].

\bibitem{erz}
 J.~R.~Ellis, G.~Ridolfi and F.~Zwirner,
  Phys.\ Lett.\ B {\bf 257} (1991) 83;
  Phys.\ Lett.\ B {\bf 262} (1991) 477;
  Yasuhiro Okada, Masahiro Yamaguchi and Tsutomu Yanagida,
{\em Phys. Lett.} B262, 54, 1991;
{\em Prog. Theor. Phys.} 85, 1, 1991;
  A.~Yamada,
  Phys.\ Lett.\ B {\bf 263}, 233 (1991);
  Howard~E. Haber and Ralf Hempfling,
{\em Phys. Rev. Lett.} 66, 1815, 1991;
M.~Drees and M.~M.~Nojiri,
  Phys.\ Rev.\ D {\bf 45} (1992) 2482;
  P.~H.~Chankowski, S.~Pokorski and J.~Rosiek,
  Phys.\ Lett.\ B {\bf 274} (1992) 191;
  Phys.\ Lett.\ B {\bf 286} (1992) 307;

  \bibitem{Degrassi:2002fi}
  G.~Degrassi, S.~Heinemeyer, W.~Hollik, P.~Slavich and G.~Weiglein,
  Eur.\ Phys.\ J.\ C {\bf 28} (2003) 133
  [arXiv:hep-ph/0212020]. 

\bibitem{estimates}
  W.~de Boer and C.~Sander,
  Phys.\ Lett.\  B {\bf 585} (2004) 276
  [arXiv:hep-ph/0307049];
  G.~Belanger, F.~Boudjema, A.~Cottrant, A.~Pukhov and A.~Semenov,
  Nucl.\ Phys.\  B {\bf 706} (2005) 411
  [arXiv:hep-ph/0407218];
  J.~R.~Ellis, K.~A.~Olive, Y.~Santoso and V.~C.~Spanos,
  Phys.\ Rev.\  D {\bf 69} (2004) 095004
  [arXiv:hep-ph/0310356];
  J.~R.~Ellis, S.~Heinemeyer, K.~A.~Olive and G.~Weiglein,
  JHEP {\bf 0502} (2005) 013
  [arXiv:hep-ph/0411216];
  J.~R.~Ellis, D.~V.~Nanopoulos, K.~A.~Olive and Y.~Santoso,
  Phys.\ Lett.\  B {\bf 633} (2006) 583
  [arXiv:hep-ph/0509331];
  J.~R.~Ellis, S.~Heinemeyer, K.~A.~Olive and G.~Weiglein,
  JHEP {\bf 0605} (2006) 005
  [arXiv:hep-ph/0602220];
  J.~Ellis, S.~Heinemeyer, K.A.~Olive, A.M.~Weber, G.~Weiglein,
  JHEP {\bf 08} (2007) 083
  [arXiv:0706.0652 [hep-ph]];
  J.~Ellis, T.~Hahn, S.~Heinemeyer, K.~A.~Olive and G.~Weiglein,
  JHEP {\bf 0710} (2007) 092
  [arXiv:0709.0098 [hep-ph]];
  J.~R.~Ellis, S.~Heinemeyer, K.~A.~Olive and G.~Weiglein,
  Phys.\ Lett.\  B {\bf 653} (2007) 292
  [arXiv:0706.0977 [hep-ph]];
  S.~Heinemeyer, X.~Miao, S.~Su and G.~Weiglein,
  JHEP {\bf 0808} (2008) 087
  [arXiv:0805.2359 [hep-ph]];
  P.~Bechtle, K.~Desch and P.~Wienemann,
  Comput.\ Phys.\ Commun.\  {\bf 174} (2006) 47
  [arXiv:hep-ph/0412012];
  R.~Lafaye, T.~Plehn, M.~Rauch and D.~Zerwas,
  Eur.\ Phys.\ J.\  C {\bf 54} (2008) 617
  [arXiv:0709.3985 [hep-ph]];
  E.~A.~Baltz and P.~Gondolo,
  JHEP {\bf 0410} (2004) 052
  [arXiv:hep-ph/0407039];
  B.~C.~Allanach and C.~G.~Lester,
  Phys.\ Rev.\  D {\bf 73} (2006) 015013
  [arXiv:hep-ph/0507283];
  B.~C.~Allanach,
  Phys.\ Lett.\  B {\bf 635} (2006) 123
  [arXiv:hep-ph/0601089];
  B.~C.~Allanach, C.~G.~Lester and A.~M.~Weber,
  JHEP {\bf 0612} (2006) 065
  [arXiv:hep-ph/0609295];
  B.~C.~Allanach and C.~G.~Lester,
  Comput.\ Phys.\ Commun.\  {\bf 179} (2008) 256
  [arXiv:0705.0486 [hep-ph]];
 B.~C.~Allanach, K.~Cranmer, C.~G.~Lester and A.~M.~Weber,
  JHEP {\bf 0708}, 023 (2007)
  [arXiv:0705.0487 [hep-ph]];
  B.~C.~Allanach and D.~Hooper,
  JHEP {\bf 0810} (2008) 071
  [arXiv:0806.1923 [hep-ph]];
  F.~Feroz, B.~C.~Allanach, M.~Hobson, S.~S.~AbdusSalam, R.~Trotta and
  A.~M.~Weber, 
  JHEP {\bf 0810} (2008) 064
  [arXiv:0807.4512 [hep-ph]];
  R.~R.~de Austri, R.~Trotta and L.~Roszkowski,
  JHEP {\bf 0605} (2006) 002
  [arXiv:hep-ph/0602028];
  L.~Roszkowski, R.~R.~de Austri and R.~Trotta,
  JHEP {\bf 0704} (2007) 084
  [arXiv:hep-ph/0611173];
  L.~Roszkowski, R.~Ruiz de Austri and R.~Trotta,
  JHEP {\bf 0707} (2007) 075
  [arXiv:0705.2012 [hep-ph]];
  L.~Roszkowski, R.~R.~de Austri, J.~Silk and R.~Trotta,
  Phys.\ Lett.\  B {\bf 671} (2009) 10
  [arXiv:0707.0622 [astro-ph]];
  R.~Trotta, F.~Feroz, M.~P.~Hobson, L.~Roszkowski and R.~Ruiz de Austri,
  JHEP {\bf 0812} (2008) 024
  [arXiv:0809.3792 [hep-ph]];
S.~S.~AbdusSalam, B.~C.~Allanach, M.~J.~Dolan, F.~Feroz and
  M.~P.~Hobson, 
  Phys.\ Rev.\  D {\bf 80} (2009) 035017
  [arXiv:0906.0957 [hep-ph]];
G.~Belanger, F.~Boudjema, A.~Pukhov and R.~K.~Singh,
  JHEP {\bf 0911} (2009) 026
  [arXiv:0906.5048 [hep-ph]]; 
S.~S.~AbdusSalam, B.~C.~Allanach, F.~Quevedo, F.~Feroz and
  M.~Hobson, 
  Phys.\ Rev.\  D {\bf 81} (2010) 095012
  [arXiv:0904.2548 [hep-ph]];
  P.~Bechtle, K.~Desch, M.~Uhlenbrock and P.~Wienemann,
  {\em Eur.\ Phys.\ J.} {\bf C 66} (2010) 215
  [arXiv:0907.2589 [hep-ph]];
  M.~Cabrera, A.~Casas and R.~Ruiz~de Austri,
  {\em JHEP} {\bf 1005} (2010) 043
  [arXiv:0911.4686 [hep-ph]];
  M.~Bridges, K.~Cranmer, F.~Feroz, M.~Hobson, R.~R.~de Austri and R.~Trotta,
  arXiv:1011.4306 [hep-ph];
  L.~Roszkowski, R.~Ruiz de Austri and R.~Trotta,
  Phys.\ Rev.\  D {\bf 82} (2010) 055003
  [arXiv:0907.0594 [hep-ph]];

\bibitem{cmssm1}
M.~Drees and M.~M.~Nojiri,
Phys.\ Rev.\ D {\bf 47} (1993) 376 [arXiv:hep-ph/9207234];
H.~Baer and M.~Brhlik,
Phys.\ Rev.\ D {\bf 53} (1996) 597 [arXiv:hep-ph/9508321];
  Phys.\ Rev.\  D {\bf 57} (1998) 567
  [arXiv:hep-ph/9706509];
  J.~R.~Ellis, T.~Falk, K.~A.~Olive and M.~Schmitt,
Phys.\ Lett.\ B {\bf 388} (1996) 97
[arXiv:hep-ph/9607292];
Phys.\ Lett.\ B {\bf 413} (1997) 355
[arXiv:hep-ph/9705444];
J.~R.~Ellis, T.~Falk, G.~Ganis, K.~A.~Olive and M.~Schmitt,
Phys.\ Rev.\ D {\bf 58} (1998) 095002
[arXiv:hep-ph/9801445];
V.~D.~Barger and C.~Kao,
Phys.\ Rev.\ D {\bf 57} (1998) 3131
[arXiv:hep-ph/9704403];
J.~R.~Ellis, T.~Falk, G.~Ganis and K.~A.~Olive,
Phys.\ Rev.\ D {\bf 62} (2000) 075010
[arXiv:hep-ph/0004169];
J.~R.~Ellis, T.~Falk, G.~Ganis, K.~A.~Olive and M.~Srednicki,
Phys.\ Lett.\ B {\bf 510} (2001) 236
[arXiv:hep-ph/0102098];
V.~D.~Barger and C.~Kao,
Phys.\ Lett.\ B {\bf 518} (2001) 117
[arXiv:hep-ph/0106189];
L.~Roszkowski, R.~Ruiz de Austri and T.~Nihei,
JHEP {\bf 0108} (2001) 024
[arXiv:hep-ph/0106334];
A.~Djouadi, M.~Drees and J.~L.~Kneur,
JHEP {\bf 0108} (2001) 055
[arXiv:hep-ph/0107316];
U.~Chattopadhyay, A.~Corsetti and P.~Nath,
Phys.\ Rev.\ D {\bf 66} (2002) 035003
[arXiv:hep-ph/0201001];
J.~R.~Ellis, K.~A.~Olive and Y.~Santoso,
New Jour.\ Phys.\  {\bf 4} (2002) 32
[arXiv:hep-ph/0202110];
H.~Baer, C.~Balazs, A.~Belyaev, J.~K.~Mizukoshi, X.~Tata and Y.~Wang,
JHEP {\bf 0207} (2002) 050
[arXiv:hep-ph/0205325];
R.~Arnowitt and B.~Dutta,
arXiv:hep-ph/0211417.

\bibitem{mc1}
O.~Buchmueller {\it et al.},
  Phys.\ Lett.\  B {\bf 657} (2007) 87
  [arXiv:0707.3447 [hep-ph]].

    \bibitem{nuhm1}
H.~Baer, A.~Mustafayev, S.~Profumo, A.~Belyaev and X.~Tata,
  Phys.\ Rev.\  D {\bf 71}, 095008 (2005)
  [arXiv:hep-ph/0412059];
            H.~Baer, A.~Mustafayev, S.~Profumo, A.~Belyaev and X.~Tata,
               {\em JHEP} {\bf 0507} (2005) 065, 
               hep-ph/0504001;
  J.~R.~Ellis, K.~A.~Olive and P.~Sandick,
  Phys.\ Rev.\  D {\bf 78}, 075012 (2008)
  [arXiv:0805.2343 [hep-ph]].

  \bibitem{mc2}
  O.~Buchmueller {\it et al.},
  JHEP {\bf 0809} (2008) 117
  [arXiv:0808.4128 [hep-ph]].

\bibitem{mc3}
  O.~Buchmueller {\it et al.},
  Eur.\ Phys.\ J.\  C {\bf 64}, 391 (2009)
  [arXiv:0907.5568 [hep-ph]].

  \bibitem{vcmssm}
  J.~R.~Ellis, K.~A.~Olive, Y.~Santoso and V.~C.~Spanos,
  Phys.\ Lett.\ B {\bf 573} (2003) 162
  [arXiv:hep-ph/0305212],
  and 
  Phys.\ Rev.\ D {\bf 70} (2004) 055005
  [arXiv:hep-ph/0405110].

\bibitem{mc4}
  O.~Buchmueller {\it et al.},
  arXiv:1011.6118 [hep-ph].

\bibitem{mc-web}
  See: {\tt http://cern.ch/mastercode}~.

\bibitem{CMSsusy}
V.~Khachatryan {\it et al.}, CMS Collaboration, arXiv:1101.1628 [hep-ex].

\bibitem{ATLASsusy}
G.~Aad {\it et al.} ATLAS Collaboration,
arXiv:1102.2357 [hep-ex].

\bibitem{BIM}
For reviews, see:
H.~P.~Nilles, Phys. Rep. {\bf 110} (1984) 1;
A.~Brignole, L.~E.~Ibanez and C.~Munoz,
arXiv:hep-ph/9707209,
published in {\it Perspectives on supersymmetry}, ed.
G.~L.~Kane, pp. 125-148;
  R.~Barbieri, S.~Ferrara and C.~A.~Savoy,
  Phys.\ Lett.\  B {\bf 119}, 343 (1982).

\bibitem{mc35}
  O.~Buchmueller {\it et al.},
  Phys.\ Rev.\  D {\bf 81} (2010) 035009
  [arXiv:0912.1036 [hep-ph]].

\bibitem{mt1733} Tevatron Electroweak Working Group and CDF Collaboration
                and D\O Collaboration,
                arXiv:1007.3178 [hep-ex].

\bibitem{btn1}
  B.~Bhattacherjee {\em et al.}, 
  arXiv:1012.1052 [hep-ph].

\bibitem{btn2}
  M.~Bona {\it et al.}  [UTfit Collaboration],
  Phys.\ Lett.\  B {\bf 687} (2010) 61
  [arXiv:0908.3470 [hep-ph]].

\bibitem{kmn}
  M.~Antonelli {\it et al.},
  Eur.\ Phys.\ J.\  C {\bf 69}, 399 (2010)
  [arXiv:1005.2323 [hep-ph]].

\bibitem{Allanach:2001kg}
  B.~C.~Allanach,
  Comput.\ Phys.\ Commun.\  {\bf 143} (2002) 305
  [arXiv:hep-ph/0104145].

\bibitem{Heinemeyer:1998np}
  S.~Heinemeyer, W.~Hollik and G.~Weiglein,
  Eur.\ Phys.\ J.\ C {\bf 9} (1999) 343
  [arXiv:hep-ph/9812472].

\bibitem{Heinemeyer:1998yj}
  S.~Heinemeyer, W.~Hollik and G.~Weiglein,
  Comput.\ Phys.\ Commun.\  {\bf 124} (2000) 76
  [arXiv:hep-ph/9812320].
  See {\tt http://www.feynhiggs.de}~.

\bibitem{Frank:2006yh}
  M.~Frank, T.~Hahn, S.~Heinemeyer, W.~Hollik, H.~Rzehak and G.~Weiglein,
  JHEP {\bf 0702} (2007) 047
  [arXiv:hep-ph/0611326].

\bibitem{Isidori:2006pk}
  G.~Isidori and P.~Paradisi,
  Phys.\ Lett.\ B {\bf 639} (2006) 499
  [arXiv:hep-ph/0605012].

\bibitem{Isidori:2007jw}
  G.~Isidori, F.~Mescia, P.~Paradisi and D.~Temes,
  Phys.\ Rev.\  D {\bf 75} (2007) 115019
  [arXiv:hep-ph/0703035], and references therein.

\bibitem{Mahmoudi:2008tp}
  F.~Mahmoudi,
  Comput.\ Phys.\ Commun.\  {\bf 178} (2008) 745
  [arXiv:0710.2067 [hep-ph]] and
  arXiv:0808.3144 [hep-ph].

\bibitem{Eriksson:2008cx}
  D.~Eriksson, F.~Mahmoudi and O.~Stal,
  JHEP {\bf 0811} (2008) 035
  [arXiv:0808.3551 [hep-ph]].

\bibitem{Heinemeyer:2006px}
  S.~Heinemeyer, W.~Hollik, D.~Stockinger, A.~M.~Weber and G.~Weiglein,
  JHEP {\bf 0608} (2006) 052
  [arXiv:hep-ph/0604147].

\bibitem{Heinemeyer:2007bw}
  S.~Heinemeyer, W.~Hollik, A.~M.~Weber and G.~Weiglein,
  JHEP {\bf 0804} (2008) 039
  [arXiv:0710.2972 [hep-ph]].

\bibitem{Belanger:2006is}
  G.~Belanger, F.~Boudjema, A.~Pukhov and A.~Semenov,
  Comput.\ Phys.\ Commun.\  {\bf 176} (2007) 367
  [arXiv:hep-ph/0607059].

\bibitem{Belanger:2001fz}
  G.~Belanger, F.~Boudjema, A.~Pukhov and A.~Semenov,
  Comput.\ Phys.\ Commun.\  {\bf 149} (2002) 103
  [arXiv:hep-ph/0112278].

\bibitem{Belanger:2004yn}
  G.~Belanger, F.~Boudjema, A.~Pukhov and A.~Semenov,
  Comput.\ Phys.\ Commun.\  {\bf 174} (2006) 577
  [arXiv:hep-ph/0405253].

\bibitem{Gondolo:2005we}
  P.~Gondolo, J.~Edsjo, P.~Ullio, L.~Bergstrom, M.~Schelke and E.~A.~Baltz,
  New Astron.\ Rev.\  {\bf 49} (2005) 149.

\bibitem{Gondolo:2004sc}
  P.~Gondolo, J.~Edsjo, P.~Ullio, L.~Bergstrom, M.~Schelke and E.~A.~Baltz,
  JCAP {\bf 0407} (2004) 008
  [arXiv:astro-ph/0406204].

\bibitem{Skands:2003cj}
  P.~Skands {\it et al.},
  JHEP {\bf 0407} (2004) 036
  [arXiv:hep-ph/0311123].

\bibitem{Allanach:2008qq}
  B.~Allanach {\it et al.},
  Comput.\ Phys.\ Commun.\  {\bf 180} (2009) 8
  [arXiv:0801.0045 [hep-ph]].

\bibitem{ATLASnote}
G.~Aad {\it et al.}, ATLAS Collaboration,
{\tt http://cdsweb.cern.ch/record/1323868/}
{\tt files/ATL-PHYS-PUB-2011-003.pdf}.

\bibitem{ATLASstudy}
G.~Aad {\it et al.}, ATLAS Collaboration,
{\tt http://cdsweb.cern.ch/record/1278474/}
{\tt files/ATL-PHYS-PUB-2010-010.pdf}.

\bibitem{Barate:2003sz}
  R.~Barate {\it et al.}  [ALEPH, DELPHI, L3, OPAL
                       Collaborations and LEP Working Group for Higgs
                       boson searches],
  Phys.\ Lett.\  B {\bf 565} (2003) 61
  [arXiv:hep-ex/0306033].

\bibitem{Schael:2006cr}
  S.~Schael {\it et al.}  [ALEPH, DELPHI, L3, OPAL
                       Collaborations and LEP Working Group for Higgs
                       boson searches],
  Eur.\ Phys.\ J.\  C {\bf 47} (2006) 547
  [arXiv:hep-ex/0602042].

\bibitem{oops}
Y.~Akrami, C.~Savage, P.~Scott, J.~Conrad and J.~Edsjo,
  arXiv:1011.4297 [hep-ph] and
  M.~Bridges, K.~Cranmer, F.~Feroz, M.~Hobson, R.~R.~de Austri and R.~Trotta, in~\cite{estimates}.

\bibitem{TeVH}
CDF and D\O\ Collaborations,\\
  arXiv:1007.4587 [hep-ex].

\bibitem{ATLASHiggs}
G.~Aad {\it et al.} ATLAS Collaboration, ATLAS--CONF-2011-005.

\bibitem{Ellis:2001qv}
  J.~R.~Ellis, S.~Heinemeyer, K.~A.~Olive and G.~Weiglein,
  Phys.\ Lett.\  B {\bf 515} (2001) 348
  [arXiv:hep-ph/0105061].

\bibitem{Ambrosanio:2001xb}
  S.~Ambrosanio, A.~Dedes, S.~Heinemeyer, S.~Su and G.~Weiglein,
  Nucl.\ Phys.\  B {\bf 624} (2002) 3
  [arXiv:hep-ph/0106255].

\bibitem{EENZ}
J.~R.~Ellis, K.~Enqvist, D.~V.~Nanopoulos and F.~Zwirner,
  Mod.\ Phys.\ Lett.\  A {\bf 1}, 57 (1986); see also
  R.~Barbieri and G.~F.~Giudice,
  Nucl.\ Phys.\  B {\bf 306} (1988) 63.

\bibitem{Strumia}
A.~Strumia,
  arXiv:1101.2195 [hep-ph].

\bibitem{LHCbbmm} 
B.~Adeva {\it et al.} [The LHCb Collaboration],
{\it Roadmap for selected key measurements of LHCb},
  arXiv:0912.4179 [hep-ex].

\bibitem{Xenon100}
E.~Aprile {\it et al.}  [XENON100 Collaboration],
  Phys.\ Rev.\ Lett.\  {\bf 105} (2010) 131302
  [arXiv:1005.0380 [astro-ph.CO]].

\bibitem{EOSavage}
J.~R.~Ellis, K.~A.~Olive and C.~Savage,
  Phys.\ Rev.\  D {\bf 77} (2008) 065026
  [arXiv:0801.3656 [hep-ph]].
 
\bibitem{Nath}
D.~Feldman, K.~Freese, P.~Nath, B.~D.~Nelson and G.~Peim,
  arXiv:1102.2548 [hep-ph].

\bibitem{Allanach}
B.~C.~Allanach,
  arXiv:1102.3149 [hep-ph].

\bibitem{Bottino}
S.~Scopel, S.~Choi, N.~Fornengo and A.~Bottino,
  arXiv:1102.4033 [hep-ph].

\end{thebibliography}
\end{document}